# Decision-Making, Sub-Additive Recursive "Matching" Noise And Biases In Risk-Weighted Stock/Bond Index Calculation Methods In Incomplete Markets With Partially Observable Multi-Attribute Preferences.

MICHAEL C. NWOGUGU
Address: Enugu 400007, Enugu State, Nigeria.
Email: Mcn2225@aol.com; mcn2225@gmail.com.
Phone: 234 814 906 2100.

**Abstract[1].**
While Indices, Index tracking funds and ETFs have grown in popularity during then last ten years, there are many structural problems inherent in Index calculation methodologies and the legal/economic structure of ETFs. These problems raise actionable issues of "Suitability" and "fraud" under US securities laws, because most Indices and ETFs are mis-leading, have substantial tracking errors and don't reflect what they are supposed to track. This article contributes to the existing literature by: a) introducing and characterizing the errors and Biases inherent in "risk-adjusted" index weighting methods and the associated adverse effects; b) showing how these biases/effects inherent in Index calculation methods reduce social welfare, and can form the basis for harmful arbitrage activities.

## 1. Existing Literature.

The literature on efficient risk-adjusted Indexing is substantial, although most of the existing research relates to the United States, Western European, Japanese, Australian and other developed capital markets (and not to emerging markets which have distinctly different characteristics). Various authors have developed alternative index weighting schemes such as risk factor benchmarks (Lee 2003; Eggins and Hill 2008; Wagner and Stocker 2009; Amenc, Goltz, Martellini & Retkowsky (2010)); and minimum variance benchmarks (Clarke, De Silva & Thorley (2006); Chan, Karceski & Lakonishok (1999)); and using firm characteristics like earnings, dividends, or book value (Arnott, Hsu & Moore (2005); Siegel, Schwartz & Siracusano (2007)); and "maximum diversification" benchmarks (Choueifaty & Coignard (2008)); and "Equal Risk Contribution" benchmarks (Maillard, Roncalli & Teiletche (2008)).

Falkenstein (2009) suggests a utility function that measures risk within the context of relative wealth and that this is an outcome of investor preference for status. This perspective is consistent with the institutional investor focus on Information Ratio as the preferred measure of risk-adjusted returns. Epaulard & Pommeret (Jan. 2001), and Backus, Routledge & Zin (December 2005); and Klibanoff, Marinacci & Mukerji (2009) have documented various classes of recursive preferences within the context of volatility and markets.

---

[1] A different version of my article was published as:
Nwogugu, M. (2013). Decision-Making, Sub-Additive Recursive "Matching Noise And Biases In Risk-Weighted Index Calculation Methods In In-Complete Markets with Partially Observable Multi-Attribute Preferences. *Discrete Mathematics, Algorithms & Applications*, 05, 1350020 (2013).

Nwogugu (2003) showed that the Put-Call Parity Theorem is inaccurate even after making adjustments for taxes and transaction costs. Wagner, Ellis & Dubofsky (1996); Ahn, Byoun & Park (2003); Hoque (Jan. 2010), and Bharadwaj & Wiggins (2001) empirically showed that the Put-Call Parity Theorem is often violated even after accounting for transaction costs. Poitras, Veld & Zabolotnyuk (2009) concluded that the Put-Call Parity Theorem is inaccurate, and that there are Early Exercise Premia for both Calls and Puts. Brunetti & Torricelli (2005) found mixed evidence about the accuracy of the Put Call Parity Theorem.

Arnott, Kalesnik, Moghtader & Scholl (Jan/Feb. 2010) compared the major index weighting methods (cap-weighting, fundamental weighting, equal weighting, minimum variance, minimum beta, risk-adjusted) and also analyzed historical returns and volatilities of such indices from 1993-2009.

Frino, Gallagher & Oetomo (2005) analyzed the daily trading and portfolio configuration strategies of index and enhanced index equity funds, and found that passive funds benefit from employing less rigid re-balancing and investment strategies; and that during index revision periods, enhanced index funds commence portfolio rebalancing earlier than index funds, and employ more patient trading strategies (all of which results in higher returns and lower trading costs for enhanced index funds). Frino, Gallagher & Oetomo (2005) also documented that where passive funds do not perfectly mimic the benchmark Index, passive funds show a greater propensity to over-weight stocks that have greater liquidity, larger market capitalization and higher past performance; and for non-index portfolio holdings, enhanced funds exhibit a greater propensity to hold 'winners' and sell 'losers'. Ang et al. (2006), and Blitz &van Vliet (2007) both found that low-volatility stocks produced higher returns while high volatility stocks produces low returns, and that the effect could not be explained by size, book-to-market, momentum and liquidity. Daniel & Titman (1997) found that firm-characteristics (rather than the covariance structure) explains the cross-sectional variation in average returns; and that when expected returns are controlled for firm characteristics, expected returns are not positively correlated to the loadings on the overall market, firm-size and book-to-market equity factors. Garlappi, Shu & Yan (2008) showed that higher default probabilities *are not* associated with higher expected stock returns; and showed that the relationship between default probability and equity return is (i) upward sloping for firms where shareholders can extract little benefit from renegotiation (low "Shareholder Advantage") and (ii) humped and downward sloping for firms with high shareholder advantage.

Haugen & Baker (1991) showed that capitalization-weighted portfolios are inefficient investments; and also defined conditions under which capitalization-weighted portfolios could be deemed the "efficient set" and showed that such conditions are not feasible. Haugen & Baker (1991) also showed that for international investors, market-matching to domestic capitalization-weighted stock indexes is likely to be a sub-optimal investment strategy when investors disagree about risk and expected return, when short-selling is restricted, when investment income is taxed, when some investment alternatives are not included in the target index, or when foreign investors are in the domestic capital market (because when these factors exist, there will be alternatives to capitalization-weighted portfolios that have the same expected return but lower volatility). Haugen & Baker (1991) also showed that that matching the market is an inefficient investment strategy - even in an informationally *efficient* market. Hsu (2006) showed that market-capitalization weighted portfolios are sub-optimal; and that cap-weighting causes a "size effect"; and over-weights stocks whose prices are high relative to their fundamentals and underweight stocks whose prices are low relative to their fundamentals; and the size of the under-performance of cap-weighted portfolios is increasing in the magnitude of price inefficiency and is roughly equal to the variance of the noise in prices. However, portfolios constructed from weights which do not depend on prices do not exhibit the same underperformance. Hsu & Campollo (Feb. 6, 2006) also showed that that market-cap weighting is not efficient.

Copeland and Zhu (2006) showed that that more sophisticated econometric models that are used to forecast Variances/Co-Variances (such as GARCH) introduce too much noise to provide cost-effective hedges. Alexander & Barbosa (2007) found that on those exchanges where Minimum-Variance hedging may still be more effective than a one-to-one hedge, it is not possible to distinguish which econometric model most efficiently reduces the variance; and that sophisticated econometric models such as GARCH introduce too much noise to provide cost-effective hedges; and that the development of both index ETFs and an advanced electronic trading networks may reduce the efficiency of a minimum-variance hedge ratio compared to a naive hedge. Juttner & Leung (2009) studied exchange rate volatilities and found that the use of high-frequency data limits the choice of the explanatory economic variables that can be included in empirical estimates; and that the first differences of GARCH(1,1) volatilities of share and bond price indices reflect portfolio trading decisions in corresponding markets for both assets; and that first differences of the gold price volatility, as an additional determinant, are related to exchange rate volatilities of two commodity currencies in the sample.

Lewellen & Nagel (2006) tested the Conditional CAPM, and showed that the variation in betas and the equity premium would have to be implausibly large to explain important asset-pricing anomalies like momentum and the value premium; and that the conditional CAPM performs nearly as poorly as the un-conditional CAPM. McLaughlin (2008) explained the evolution and structure of ETFs, and documented some problems inherent in Indices and ETFs such as Index Arbitrage and ETF arbitrage and suggested some solutions. Chen, et. al. (2006) also documented losses attributable to Index-arbitrage and ETF-arbitrage (which have not been solved by current index weighting methods). Walsh (1997) compared the usefulness of trades versus Orders, and found that both permanent and temporary price effects are order-size-related; and orders are clearly better measures than trades; and the ability to measure information asymmetry in order flow, increases with trading volume.

Mar, Bird, Casavecchia & Yeung (2009) examined fundamental indexation in an Australian context over the period 1995 to 2006 and found that the superiority of fundamental indexation (over cap-weighted indexation) is largely explained by its inherent bias towards value stocks, which raises the question as to whether a more overt value tilt may not provide a superior means for exploiting mis-pricings in markets. Fernholz, et al. (1998) developed a "stock market diversity" measure (distribution of capital in an equity market first introduced in Fernholz (1999)), as a weighting factor. Under somewhat simplifying assumptions, Fernholz (1999) shows that the return of a *Diversity-Weighting* scheme relative to the market-capitalization weighting scheme is, among other things, a function of the difference between the weighted-average variance of the individual stocks and the portfolio variance. Choueifaty & Coignard (2006) developed a portfolio optimization method named the "*Diversification Index*", which was defined as the ratio of average volatility of the stocks in the portfolio divided by the portfolio volatility. DeMiguel et al. (2007) found that estimation errors in input parameters almost entirely invalidates the performance of formal optimization models, even when improved estimators are used, unless for unreasonably large sample size. Zitzewitz (2003) documented the evolution and scope of the Index arbitrage and mutual fund arbitrage problem and the rather slow responses of mutual funds; and found that the speed and efficacy of funds' reactions is negatively correlated with expense ratios and the share of insiders on the board, suggesting that fund governance may be important in determining whether funds take actions to protect their shareholders. Drew, Naughton & Veeraraghavan (2003) tested the multifactor approach to asset pricing in the Shanghai Stock Exchange and also tested for the size and value premia, and found that mean-variance efficient investors in that market can select some combination of small and low book-to-market equity firms in addition to the market portfolio to generate superior risk-adjusted returns; and that seasonal effects dont explain the findings of the multifactor model, and that the market factor alone is not sufficient to describe the cross-section of average stock returns in China; and the book-to-market equity effect is not as pervasive as was found for the United States portfolios.

Index calculation methods from other fields provide useful insights on common problems in representativeness and accuracy of indices – relevant articles include Nayebi (2006); Murphy & Garvey (2005); Schultz (2001); Bernhard (1971); Jha, Murthy & Bhanu (April 2003); U.S. Department Of Commerce/National Oceanic and Atmospheric Administration (2003); Hurlbert (1971); Von Der Lippe (1999); Aydin & Ozer (2005); Hertzberg (1987); Neher & Darby (_____); Ramsden (2009); Berger & Pukthuanthong (2012); Caves, Christensen & Diewert (1982); Diewert E (2009); and Karabatsos (2000). Forthman, Gold, Dove & Henderson (2010), DesHarnais, Forthman, Homa-Lowry & Wooster (2000), and DesHarnais, Forthman, Homa-Lowry & Wooster (1997), discussed the use of indices for measuring the quality of healthcare.

For alternative index construction approaches that are partly based on, or rely on decision-making models, see Jianping, et. al. (2012); Chen (2012); Li (2009); Bi, Ding, Luo & Liang (2011); and Su (2011). On the dispersion of opinions and valuations in financial markets, see Takahashi (2012).

The specific gaps in the existing literature include the following:

a) Analysis of biases and behavioral effects that are inherent in, or are caused by different index weighting methods.
b) The effects of the Put-Call Parity Theorem on Index weighting methods.
c) The effects of static Index Revision Dates on Index weighting methods and Index arbitrage.
d) Non-legislative methods for eliminating Index arbitrage.

e) The effects of the popularity of valuation ratios and accounting data on efficiency of indices and correlations.
f) The effects of the validity of ICAPM on index weighting methods.

This article does not cover options-based indices (which are addressed in Nwogugu (2010a)) or Leveraged/Inverse ETFs (which are addressed in Nwogugu (2010b) or Credit-Default-Swap Indices (which are addressed in Nwogugu (2010c). The rest of this article analyzes various traditional index calculation methods and explains the inherent short-comings and biases.

## 2. The ICAPM/CAPM Is Inaccurate

Nwogugu (2003); Nwogugu (2010e), Prono (Jan. 2010); Prono (June 2007); Green & Hollifield (1992); Guo (May/June 2004); Kumar & Ziemba (1993); Lewellen & Nagel (2006); Drew, Naughton & Veeraraghavan (2003); Gharghori, Chan & Faff (2007); Tofallis (2008) have shown that ICAPM and CAPM are inaccurate.

***Theorem-1: For Any Investment Horizon And Any Market, Risk-Weighted Index Weighting Methods Don't Reflect The Market Portfolio*.**

***Proof***: ICAPM/CAPM and Variance/Correlation/Covariance are inaccurate and thus, traditional index weighting methods (capitalization weighting; Fundamental weighting; Trading Volume Weighting; Equal Weighting) are not accurate because the "Market portfolio" or Risk-weighted portfolios are not the most efficient portfolio in terms of risk-reward trade-offs. Thus, contrary to finance theory, under a "standard" interpretation of the ICAPM/CAPM, the Sharpe Ratio of a broad-based risk-weighted portfolio (a "market" portfolio) is not automatically maximized (there is no mean-variance optimality). Mean-Variance Risk-weighted Indices are not "efficient indices" because they don't reflect the "market portfolio" and the true opportunity set available to investors; because they don't contain all the shares/companies in the market and don't contain and all possible strategies (including short positions, long positions and margin positions), and hence do not truly represent the market. The ideal "Market portfolio" must contain all Stocks and strategies in the market. Furthermore, while Risk-weighted Indices are "long-only positions", investors can short securities, or buy securities on margin; or buy non-listed securities. Risk-Weighting methodologies erroneously assume that all investors have the same preferences, and derive the same utilities/dis-utilities from both final end-of-period asset prices and intermittent changes in asset prices. □

***Theorem-2: For Any Investment Horizon And Any Market, All Risk-weighting Methods Distort The Risk Of Constituent Companies*.**

***Proof***: Risk weighting methodologies distort the true state of the asset market because in reality, the price-risk relationship for any asset and the mean of the asset's prices are not always constant over any or all periods of time, and are not constant for any group of investors, and are not constant for any equal or unequal successive changes in prices of the asset. Constant price-risk relationship and constant mean are major assumptions of most risk-weighting methods. Secondly, the performance of an asset in its natural trading state differs substantially from the risk-based weight assigned to the asset in the risk weighted index. This because investors' evaluation of perceived and actual risk often depend on relative prices of assets. □

***Theorem-3: The Formula For Covariance Is Inaccurate.***

***Proof***: The general formulas for Correlation, Covariance and Variance are as follows:

Covariance = $C_{cov} = [\{(\chi_1-\mu_x)*(\gamma_1-\mu_y)\}+\{(\chi_2-\mu_x)*(\gamma_2-\mu_y)\}.+..+\ldots\ldots(\chi_n-\mu_x)*(\gamma_n-\mu_y)]/\eta$.

Correlation = $C_{cor} = C_{cov}/(\sigma_y*\sigma_x)$

Variance = $V = \sigma_x^2 = [(\chi_1-\mu_x)^2+(\chi_2-\mu_x)^2.+..+\ldots\ldots(\chi_n-\mu_x)^2]/\eta$.

Semi-variance = $SV = [Min\{(\chi_1-\mu_x),0\}^2+Min\{(\chi_2-\mu_x),0\}^2.+..+\ldots\ldots Min\{(\chi_n-\mu_x),0\}^2]/\eta$.

Where:
$\boldsymbol{\mu_{x\eta}}$ = mean of variable $\boldsymbol{\chi}$ for $\eta$ periods.

$\boldsymbol{\mu}_{\mathbf{y}\eta}$ = mean of variable $\boldsymbol{\gamma}$, for $\eta$ periods.
$\boldsymbol{\chi}_{\mathbf{i}}$ = value of variable $\boldsymbol{\chi}$ in period i.
$\boldsymbol{\gamma}_{\mathbf{i}}$ = value of variable $\boldsymbol{\gamma}$ in period i.
$\mathbf{C}_{\mathbf{cor}}$ = correlation.
**N** = number of periods. $\eta \in N$, where $\eta$ is the number of time intervals. $i \in \eta$.
$\mathbf{C}_{\mathbf{cov}}$ = covariance.
$\boldsymbol{\sigma}_{\mathbf{x}\eta}$ = standard deviation of x in period $\eta$.
$\boldsymbol{\sigma}_{\mathbf{y}\eta}$ = standard deviation of y in period $\eta$.
**V** = variance
**SV =** Semivariance
$\mathbf{U}_{\mathbf{x}}^{+}/\mathbf{U}_{\mathbf{x}}^{-}$ = utility/disutility gained from increase/decrease in variable x.
U = utility/disutility from covariance/co-movement of two series.
$\Psi$ = Skewness
$\kappa$ = Kurtosis
t = tax
r = interest rate
Users sometimes make minor modifications to these formula where large data sets are used (eg. number of trading days in a year, when analyzing stock prices).
The current Covariance formulas can be valid <u>iff</u> <u>all</u> the following conditions exist simultaneously:

**1.** $\partial^3\kappa/\partial\chi_1\partial\eta\partial\boldsymbol{\sigma}_{\mathbf{x}\eta} = 0$
**2.** $\partial V/\partial\eta = 0$; $\partial^2 V/\partial\eta\partial\boldsymbol{\mu}_{\mathbf{x}\eta} = 0$; $\partial^3\boldsymbol{\sigma}_{\mathbf{x}\eta}/\partial V\partial\eta\partial\boldsymbol{\mu}_{\mathbf{x}\eta} = 0$
3. $\partial V/\partial\boldsymbol{\mu}_{\mathbf{x}\eta} = 0$; $\partial V/\partial\boldsymbol{\mu}_{\mathbf{y}\eta} = 0$
**4.** $\partial[(\chi_1-\mu_{x\eta})^2]/\partial\eta = 0$
**5.** $\partial\sigma_{x\eta}/\partial\eta = 0$;
**6.** $\partial\sigma_{x\eta}/\partial[(\chi_1-\mu_{x\eta})^2] = 1$;
**7.** $\partial[(\chi_1-\mu_{x\eta})^2]/\partial\boldsymbol{\mu}_{\mathbf{x}\eta} > 0$; this implies that Variance/Covariance/Correlation are not valid for any series or pair of series, in which most of the numbers are less than one.
**8.** $\partial^2 C_{cov}/\partial\mu_{x\eta}\partial\mu_{y\eta} = 0$;
**9.** $\partial[(\chi_1-\mu_{x\eta})]/\partial[(\gamma_i-\mu_{y\eta})] = 0$;
**10.** $\partial\mu_{x\eta}/\partial\mu_{y\eta} = 0$;
**11.** $(\partial\chi_1/\partial\gamma_i)*(\partial\mu_{x\eta}/\partial\mu_{y\eta}) > 0$;
**12.** $\partial\chi_1/\partial\gamma_i \leq \partial\mu_{x\eta}/\partial\mu_{y\eta}$
13. $\partial^2 V/\partial\eta\partial\boldsymbol{\sigma}_{\mathbf{x}\eta} = 1$;
14. $\partial(\chi_i-\mu_{x\eta})/\partial\eta = 0$;
15. $\partial\boldsymbol{\sigma}_{\mathbf{x}\eta}/\partial[(\chi_1-\mu_{x\eta})] \geq 1$;
16. $\partial\mu_{x\eta}/\partial\eta = 0$; and $\partial^2\mu_{x\eta}/\partial\eta^2 = 0$;
17. $\partial[(\chi_1-\mu_x)]/\partial\mu_{x\eta} = 0$; and $\partial^2[(\chi_1-\mu_{x\eta})^2]/\partial\mu_{x\eta}^2$
18. $\partial\chi_1/\mu_{x\eta} = 0$; and $\partial^2\chi_1/\mu_{x\eta}^2 = 0$;
19. $\partial\Psi/\partial\mu_{x\eta} = 0$; and $\partial^2\Psi/\partial\mu_{x\eta}^2 = 0$;
20. $\partial^3\Psi/\partial\eta\partial\chi_1\partial\mu_{x\eta} = 0$;
**21.** $\boldsymbol{\partial^2\kappa/\partial(\chi_1-\mu_{x\eta})\partial\eta > 1}$**;**
22. $\partial\kappa/\partial\Psi \geq 1$; and $\partial^2\kappa/\partial\Psi^2 > 0$;
23. $\partial\kappa/\partial\chi_1 > 0$; and $\partial^2\kappa/\partial\chi_1^2 > 0$;
24. $\partial^2\Psi/\partial\chi_1\partial\eta = 0$;
25. $\partial V/\partial\boldsymbol{\sigma}_{\mathbf{x}\eta} = 1$; $\partial^2 V/\partial\boldsymbol{\sigma}_{\mathbf{y}\eta}^2 > 0$;
26. $\partial(\chi_i-\mu_{x\eta})/\partial\eta = 0$; and $\partial(\boldsymbol{\gamma}_{\mathbf{i}}-\mu_{y\eta})/\partial\eta = 0$;
27. $\partial\mu_{x\eta}/\partial\chi_i > 0$; $\partial\mu_{y\eta}/\partial\boldsymbol{\gamma}_{\mathbf{i}} = 0$;
28. $\partial\chi_1/\partial\mu_{y\eta} = 0$; $\partial\gamma_i/\partial\mu_{x\eta} = 0$; this implies that any significant positive or negative co-movement between the mean of one series and the other series renders covariance/variance/semi-variance useless.
**29.** $\partial^2 U/\partial\boldsymbol{\sigma}_{\mathbf{x}\eta}\partial\boldsymbol{\mu}_{\mathbf{x}\eta} = 0$ (finance only)
30. $\partial^2 U/\partial\Psi\partial\kappa > 0$;
31. $\partial\boldsymbol{\chi}_{\mathbf{i}}/\partial\boldsymbol{\chi}_{\mathbf{(i-1)}}) = 0$; and $\partial\gamma_{\mathbf{i}}/\partial\boldsymbol{\gamma}_{\mathbf{(i-1)}}) = 0$;

32. $\partial U/\partial V = 0$ (finance only)
33. $\partial^2 U/\partial\eta\partial t = 0$ (finance only)
34. $\partial^2 C_{cov}/\partial\eta\partial r = 0$ (finance only)
35. $\partial^3 U/\partial\eta\partial\Psi\partial\boldsymbol{\sigma}_{x\eta} = 0$ (finance only)
36. $\partial^2 U/\partial\boldsymbol{\chi}_i\partial r > 0$ (finance only)
37. $\partial(\chi_i-\mu_{x\eta})/\partial r = 0$ (finance only)
38. $\partial[(\chi_i-\mu_{x\eta})^2]/\partial\chi_i = 1$; $\partial^2[(\chi_i-\mu_{x\eta})^2]/\partial\chi_i^2 > 0$;
39. $\partial U/\partial(\chi_i-\mu_{x\eta}) > 0$; and $\partial^2 U/\partial(\chi_i-\mu_{x\eta})^2 > 0$; (finance only)

Since <u>none</u> of these conditions are feasible, Covariance is inaccurate.□

***<u>Theorem 4: The Variance And Semi-Variance Formulas Are Wrong Under All Circumstances.</u>***

***Proof***: The current Variance and Semi-Variance formulas can be valid <u>iff</u> <u>all</u> the following conditions exist:

**1.** $\partial^3 K/\partial\chi_1\partial\eta\partial\boldsymbol{\sigma}_{x\eta} = 0$
**2.** $\partial V/\partial\eta = 0$; $\partial^2 V/\partial\eta\partial\mu_{x\eta} = 0$;
**3.** $\partial V/\partial\boldsymbol{\mu}_{x\eta} = 0$;
**4**. $\partial[(\chi_1-\mu_{x\eta})^2]/\partial\eta = 0$ ???
**5**. $\partial\sigma_{x\eta}/\partial\eta = 0$;
**6**. $\partial\sigma_{x\eta}/\partial[(\chi_1-\mu_{x\eta})^2] = 1$
**7**. $\partial[(\chi_1-\mu_x)^2]/\partial\boldsymbol{\mu}_{x\eta} > 0$; this implies that Variance/Covariance/Correlation are not valid for any series or pair of series in which most of the numbers are less than one, and or the means are between zero and one.
**8**. $\partial^2 C_{cov}/\partial\mu_{x\eta}\partial\mu_y = 0$
**9**. $\partial[(\chi_1-\mu_{x\eta})]/\partial[(\gamma_i-\mu_{y\eta})] = 0$
**10.** $\partial\mu_{x\eta}/\partial\mu_{y\eta} = 0$
**11**. $[(\partial\chi_1/\partial\gamma_i)*(\partial\mu_{x\eta}/\partial\mu_{y\eta})] > 0$; and $[(\partial^2\chi_1/\partial\boldsymbol{\gamma}_i^2)*(\partial^2\mu_{x\eta}/\partial\mu_{y\eta}^2)] > 0$;
**12**. $\partial\chi_1/\partial\gamma_i \leq \partial\mu_{x\eta}/\partial\mu_{y\eta}$
**13**. $\partial^2 V/\partial\eta\partial\boldsymbol{\sigma}_{x\eta} = 1$
**14**. $\partial(\chi_i-\mu_{x\eta})/\partial\eta = 0$;
**15**. $\partial\boldsymbol{\sigma}_{x\eta}/\partial[(\chi_1-\mu_x)] \geq 1$
**16**. $\partial\mu_{x\eta}/\partial\eta = 0$; and $\partial^2\mu_{x\eta}/\partial\eta^2 = 0$;
**17**. $\partial[(\chi_1-\mu_{x\eta})]/\partial\mu_{x\eta} = 0$; ?????
**18**. $\partial\chi_1/\partial\mu_{x\eta} = 0$; and $\partial^2\chi_1/\mu_{x\eta}^2 = 0$;
**19**. $\partial\Psi/\partial\mu_{x\eta\eta} = 0$; and $\partial^2\Psi/\partial\mu_{x\eta}^2 = 0$
**20**. $\partial^3\Psi/\partial\eta\partial\chi_1\partial\mu_{x\eta} = 0$
**21. $\boldsymbol{\partial^2\kappa/\partial(\chi_1-\mu_{x\eta})\partial\eta > 1}$;**
**22.** $\partial\kappa/\partial\Psi \geq 1$; and $\partial^2\kappa/\partial\Psi^2 > 0$
**23.** $\partial\kappa/\partial\chi_1 > 0$; and $\partial^2\kappa/\partial\chi_1^2 > 0$;
**24.** $\partial^2\Psi/\partial\chi_1\partial\eta = 0$
**25.** $\partial V/\partial\boldsymbol{\sigma}_{x\eta} = 1$; $\partial^2 V/\partial\boldsymbol{\sigma}_{x\eta}^2 > 0$
**26.** $\partial(\chi_i-\mu_{x\eta})/\partial\eta = 0$;
**27**. $\partial^2 U/\partial\boldsymbol{\sigma}_{x\eta}\partial\boldsymbol{\mu}_{x\eta} = 0$ (finance only)
**28**. $\partial^2 U/\partial\Psi\partial\kappa > 0$;
**29**. $\partial\boldsymbol{\chi}_i/\partial\boldsymbol{\chi}_{(i-1)} = 0$; and $\partial\boldsymbol{\gamma}_i/\partial\boldsymbol{\gamma}_{(i-1)} = 0$;
**30**. $\partial U/\partial V = 0$ (finance only)
**31**. $\partial^2 U/\partial\eta\partial t = 0$ (finance only)
**32**. $\partial^2 C_{cov}/\partial\eta\partial r = 0$ (finance only)
**33**. $\partial^3 U/\partial\eta\partial\Psi\partial\boldsymbol{\sigma}_{x\eta} = 0$ (finance only)
**34**. $\partial^2 U/\partial\boldsymbol{\chi}_i\partial r > 0$ (finance only)
**35**. $\partial(\chi_i-\mu_{x\eta})/\partial r = 0$ (finance only)
**36**. $\partial[(\chi_i-\mu_{x\eta})^2]/\partial\chi_i = 1$; $\partial^2[(\chi_i-\mu_{x\eta})^2]/\partial\chi_i^2 > 0$;
**37.** $\partial U/\partial(\chi_i-\mu_{x\eta}) > 0$; and $\partial^2 U/\partial(\chi_i-\mu_{x\eta})^2 > 0$; (finance only)
**38.** $\partial SV/\partial\boldsymbol{\sigma}_{x\eta} = 1$; $\partial^2 V/\partial\boldsymbol{\sigma}_{x\eta}^2 > 0$
**39**. $\partial U/\partial SV = 0$ (finance only)

**40.** $\partial SV/\partial \eta = 0$; $\partial^2 SV/\partial \eta \partial \mu_{x\eta} = 0$;
**41.** $\partial SV/\partial \boldsymbol{\mu}_{\mathbf{x}\eta} = 0$;
**42**. $\partial^2 SV/\partial \eta \partial \boldsymbol{\sigma}_{\mathbf{x}\eta} = 1$

Since none of these conditions are feasible, Variance and Semi-Variance are inaccurate. □

***Theorem-6: The Sharpe Ratio, the Treynor Ratio and the Jensen Alpha Are Inaccurate.***
***Proof***: See: Danielsson, Jorgensen & Sarma & Vries (2006); Nwogugu (2005a;2005b); Broadie, Chernov & Johannes (2009). Haugen & Baker (_____). Taleb (2008). The Sharpe Ratio, the Treynor Ratio and the Jensen Alpha may be accurate and or efficient risk measures *iff* all the following assumptions exist simultaneously:

1) All investors agree about the risk and expected return for all securities.
2) All investors can short-sell all securities without restriction.
3) All investors have the same or similar investment horizon, or their investment decisions dont consider investment horizons.
4) All investors dont pay federal or state income taxes.
5) There are minimal or no Transaction costs.
6) The investment opportunity set for all investors holding any security in the index is restricted to the securities in the public markets (or in the specific sub-market on which the Index is based).
7) All Investors have similar decision processes about investments.
8) For every investor, the risk-reward trade-off is a more important investment criteria than the absolute magnitude of returns, and or the investor's "reference point" (ie. Cost of capital; etc.).
9) Bid-ask spreads are small and dont affect investors' decisions or the calculation of standard deviations of returns.
10) All losses produce strictly negative utilities (investors dont gain any utility from tax loss carry-forwards).
11) For all investors, the non-monetary utilities (such as hedging, long term security, etc.) that arise from investing are irrelevant.
12) All investors can make investments that earn the risk-free rate at all times and for any amount of capital.
13) The returns of the underlying asset have a normal distribution.
14) The rate of change of the standard deviation with respect to the realized return ($\partial\sigma/\partial r$) is constant for any time interval during the investment horizon.
15) The realized return (r) of the underlying asset is constant over any time interval during the investment horizon; and as *t* (time) $\rightarrow \infty$, $\partial\sigma/\partial r$ is constant.
16) As $r \rightarrow \infty$, $\partial\sigma/\partial r$ is constant.
17) Any correlation between $\sigma$ and r, or between $\partial\sigma$ and $\partial r$, is irrelevant.
18) ICAPM/CAPM are valid.

Given that these assumptions cannot all be feasible simultaneously, the Sharpe Ratio, the Treynor Ratio and the Jensen Alpha are not efficient or accurate risk measures, and are not efficient basis for any index weighting system. As explained in Nwogugu (2003; 2010b), CAPM and ICAPM are wrong. Furthermore, the risk-adjusted performance measures developed by Treynor (1965), Sharpe (1966) and Jensen (1968) were all based on cap-weighted indexes which were used in all of these studies as the benchmark for measuring the performance of active investment managers – but as explained in Nwogugu (2010c) and in Haugen & Baker (1991), cap-weighted indices are inaccurate and inefficient. The Mean-Variance framework is highly sensitive to the biases inherent in computed returns, some of which are explained in Nwogugu (2010d). Conrad & Kaul (1993). □

## 3. The Risk-Adjusted Index Calculation Methods Are Wrong.

### 3.1 Free Float Adjusted Indices.

Free Float Adjusted Indices suffer from all of the weaknesses of Market-cap weighted Indices. Free-float Adjustment is error because it does not achieve the intended objective of reflecting the true "opportunity set" available to investors that want to invest in a subject company. This is because: a) there can be PIPE transactions and private placements through which public investors can purchase the company's shares, b) the subject non-traded Shares of the company are held by "investors" who are typically not different from traditional investors, c) investors can also take positions in the company through its publicly-traded Put/Call Options, d) the subject company's publicly traded convertible bonds and convertible preferred stock are also alternative ways to take positions in the company, but the free Float Adjustment does not consider these securities; e) where the float is relatively small (less than sixty Percent of total shares outstanding), Free Float Adjustment significantly distorts the contribution of the company to the index and vastly under-weights the company in the Index, f) the subject company's publicly traded bonds are also another way of investing in the company any Free Float Adjustment should also include adjustments for such bonds.

### 3.2 Equal Risk Contribution ("ERC") Indices

These are a hybrid of equal-weighting methods and risk-contribution methods, and thus, have the dis-advantages of both methods. Neurich (2008); Taleb (2008). Qian (2005; 2006); Maillard, Roncalli & Teiletche (2008). In ERC methods, risk is measured solely in terms of standard deviations which introduces market noise and other factors. Taleb (2008), Nwogugu (2005) and Nwogugu (2008) explain why standard deviations and Variances are not good risk measures; and the proofs above also explain why variance is an inaccurate measure of risk.

The ERC index calculation method often does not consider the effect of correlations among stocks/assets selected for inclusion in an index. The ERC method causes increased contagion and correlation. This is because the risk contribution is measured in terms of standard deviation and variances. Hence, if the standard deviation of many companys' stock prices are highly positively correlated with that of the "market", contagion will be high in up-markets, and if the standard deviation of many companys' stock prices are negatively correlated with that of the "market", contagion will be high in down-markets. The ERC method causes excessive re-balancing costs. In order for the Index to be accurate, the ERC methods require instantaneous and continuous Index revision/re-balancing. The logic of the ERC method is flawed because the risk contribution of each stock/asset is partly dependent on its correlation with other stocks/assets in the index and its volatility, all of which change instantaneously.

ICAPM/CAPM and Variance/Correlation/Covariance are inaccurate and thus, "Diversity" weighting methods are not accurate because the "Market portfolio" or the "ERC" portfolio are not the most efficient portfolio in terms of risk-reward trade-offs. Nwogugu (2003); Prono (Jan. 2010; June 2007); Green & Hollifield (1992); Guo (May/June 2004); Kumar & Ziemba (1993); Lewellen & Nagel (2006). The ERC weighting method doesn't reflect the "market portfolio" and the true opportunity set available to investors; because: i) the variances of returns of each company's shares change instantaneously and no manager has real time access to the information for all public companies, ii) the ERC portfolio doesn't contain all the shares/companies in the market and doesn't contain and all possible strategies, iii) the true opportunity set available to investors are not completely defined by risk alone, iv) investor have different preferences about risk, and minimum-variance my not bean investment objective for all investors, v) in some circumstances, losses have utilities, and or are transferable, vi

Most ERC methodologies revise the Index weight periodically and on specific dates, and this increases the Index tracking error because: a) most changes in the prices of the underlying Shares occur during the index revision period and are not captured by the index weights, and b) because arbitrageurs know the exact dates of the index re-balancing, they distort prices of shares of constituent companies around index re-balancing dates, c) index funds and Index ETFs must re-balance on the designated days, and thus have weak negotiating positions and are forced to become price-takers, which in turn, distorts the underlying Index. These factors do not enhance price discovery. The combination of the ERC methods and the static index revision dates substantially reduces the Index's tracing accuracy.

ERC Index calculation methods cause a *Volatility Bias* (over-weights securities that have low-volatility and under-weights securities that have high volatility); and a *Correlation Bias* (over-weights securities that have a high positive or negative correlation with the "market" or major stock indices; and vice-versa). ERC Index calculation methods cause: a) a *Volume Bias* (where Volatility is positively correlated to daily trading volume, these Index methods under-weights securities that are frequently traded and over-weights securities that have high daily trading volumes); and b) an *Optimization-Rule Bias* (over-weights or under-weights securities depending on whether there are weight constraints/caps in the optimization process); c) a *Substitution Bias* (the probability of substitution of a security in the index is directly proportional to its Correlation to the rest of the portfolio); and d) a *Noise Indifference Effect* (ERC Index calculation don't consider the amount of "noise" in security prices; and thus, will under-weight securities whose prices contain significant noise in periods of high-volatility; and will overweight securities whose prices contain significant noise in periods of low volatility.

**3.3 "Most-Diversified" ("Diversity") Indices.**

These "Diversification" indices are a hybrid of equal-weighting methods and risk-contribution methods. Choueifaty & Coignard (2008); Taleb (2008); Choueifaty (2010); Fernholz, Garvy & Hannon (1998). An example is the Anti-benchmark portfolios (AB) introduced in Choueifaty & Coignard (2008). The construction of these indices is based almost entirely on Covariances/Variances – there are no forecasts of expected returns of the constituent companies or other factors. The "Diversity" Indices are highly sensitive to errors in Covariances and variances – both of which have been shown herein to be inaccurate measures of risk. Kumar & Ziemba (1993). There is theoretical and empirical evidence that risk preferences vary among individual investors; and that the classic definition of risk as the volatility of total return is inconsistent with investor experience and market trends. Alexander & Barbosa (2007) found sophisticated econometric models such as GARCH introduce too much noise to provide cost-effective hedges; and that the development of advanced electronic trading networks may reduce the efficiency of a minimum-variance hedge ratio compared to a naive hedge. The "Diversity" Index calculation methods cause the following Biases: **i)** a *Volatility Bias* (over-weights securities that have low-volatility and under-weights securities that have high volatility); **ii)** A ***Correlation Bias*** (overweighs securities that have a high positive or negative correlation with the "market" or major stock indices; and vice-versa); **iii)** A *Volume Bias* (where Volatility is positively correlated to daily trading volume, these Index methods under-weights securities that are frequently traded and over-weights securities that have high daily trading volumes); **iv)** An *Optimization-Rule Bias* (over-weights or under-weights securities depending on whether there are weight constraints/caps in the optimization process); **v)** A *Dividend Bias* (under-weights dividend-paying securities and over-weights non-dividend-paying securities – Dividends are not included in historical or realized volatility estimates, and dividend-paying securities tend to be less volatile than non-dividend-paying securities); **vi)** A *Substitution Bias* (the probability of substitution of a security in the index is directly proportional to its Correlation to the rest of the portfolio); **vii**) A *Noise Indifference Effect* (these methods don't consider the amount of noise in security prices; and thus, will under-weight securities whose prices contain significant noise in periods of high-volatility; and will overweight securities whose prices contain significant noise in periods of low volatility; **viii)** A *Volatility Estimation Bias* (over-weights or under-weights securities depending on the method used to calculated historical volatility or realized volatility or forecasted volatility); and **ix)** *Selection/Risk-Reduction Bias* (because the main criteria is the impact of addition of a security on an existing portfolio, the probability of inclusion in the index heavily depends on the nature of already-selected securities).

As described in Nwogugu (2005a;b), standard deviation is not a good measure of risk. Furthermore, optimization processes are often dependent on assumed probability distributions. Secondly the volatility-based "diversification" is relies entirely on market risk, increases correlations, and hence incorporates substantial market noise and is removed from the fundamental operations and operational risk of the constituent companies. As described above, continued focus on market risk and volatility is a vicious circle that fosters and creates more correlation and volatility.

ICAPM/CAPM and Variance/Correlation/Covariance are inaccurate and thus, "Diversity" weighting methods are not accurate because the "Market portfolio" or the "Diversity" portfolio (the most-diversified portfolio under mean-variance framework) are not the most efficient portfolio in terms of risk-reward trade-offs. Nwogugu (2003); Nwogugu (2007); Nwogugu (2005). Prono (Jan. 2010); Prono (June 2007); Green & Hollifield (1992); Guo (May/June 2004); Kumar & Ziemba (1993); Lewellen & Nagel (2006); Drew, Naughton & Veeraraghavan (2003). The "Diversity" weighting method doesn't reflect the "market portfolio" and the true opportunity set available to investors; because of the following reasons: **i)** the variances of returns of each company's shares change instantaneously and no manager has real time access to the information for all public companies, **ii)** the "Diversity" portfolio doesn't contain all the shares/companies in the market and don't contain and all possible strategies, **iii)** the true opportunity set available to investors are not completely defined by risk alone, **iv)** investor have different preferences about risk, and minimum-variance my not bean investment objective for all investors, **v)** in some circumstances, losses have utilities, and or are transferable.

**3.4 "Minimum-Variance" Indices.**

Minimum variance indices are constructed by minimizing the volatility of constituent securities without reference to return expectations. Taleb (2008). Alexander & Barbosa (2007). Scherer (Sept. 2011). Amenc, Goltz, Martellini & Ye (2011). Haugen and Baker (1991) found that due to investor restrictions on short selling, tax situations, and risk and return expectations, portfolios could be constructed that dominated the market portfolio in terms of risk-adjusted returns. The Minimum Variance Indices are highly sensitive to errors in estimates of Covariances and variances. The results in Daniel & Titman (1997) and Copeland & Zhu (2006) imply that minimum-Variance Indices are inaccurate and don't reflect the true risk of the subject companies – and contradict many findings and conclusions in Amenc, Goltz, Martellini & Ye (2011). Kumar & Ziemba (1993). The *Minimum-Variance* index calculation methods cause: i) a *Volatility Bias* (over-weights securities that have low-volatility and under-weights securities that have high volatility); ii) a *Correlation Bias* (over-weights securities that have a high positive or negative correlation with the "market" or major stock indices; and vice-versa); iii) a *Volume Bias* (where Volatility is positively correlated to daily trading volume, these Index methods under-weights securities that are frequently traded and over-weights securities that have high daily trading volumes); **iv)** an *Optimization-Rule Bias* (over-weights or under-weights securities depending on whether there are weight constraints/caps in the optimization process); **v)** a *Substitution Bias* (the probability of substitution of a security in the index is directly proportional to its Correlation to the rest of the portfolio); **vi)** a *Noise Indifference Effect* (these methods don't consider the amount of noise in security prices; and thus, will under-weight securities whose prices contain significant noise in periods of high-volatility; and will overweight securities whose prices contain significant noise in periods of low volatility; and **vii)** a *Volatility Estimation Bias* (over-weights or under-weights securities depending on the method used to calculated historical volatility or realized volatility or forecasted volatility); and **viii)** *Selection/Risk-Reduction Bias* (because the main criteria is the impact of addition of a security on an existing portfolio, the probability of inclusion in the index heavily depends on the nature of already-selected securities).

An example of Minimum-Variance Index is the MSCI Barra's "MSCI Global Minimum Volatility Indices" which focus on absolute return and low volatility. The MSCI Barra indices are constructed by performing total risk minimizing optimization using MSCI parent indices and the Barra Global Equity Model (GEM2) as the risk estimate input. MSCI Barra currently calculates six MSCI Minimum Volatility Indices: for USA, Europe, World, EAFE, Emerging Markets and "All Country World". See: MSCI Global Minimum Volatility Indices Methodology Guide (November 2009). http://www.mscibarra.com/products/indices/thematic_and_strategy/minimum_volatility/MSCI_Minimum_Volatility_Methodology.pdf.

However, there is empirical evidence that risk preferences vary among individual investors; and that the classic definition of risk as the volatility of total return is inconsistent with investor experience and market trends; and is often far removed from daily trends in the real economy. As described in Nwogugu (2005a;b) and Nwogugu (2010b), standard deviation and correlation are not good measures of risk. Secondly, the volatility-based "diversification" is relies entirely on market risk, increases correlations among assets in markets, and hence incorporates substantial market noise and is very much removed from the fundamental operations and operational risk of the constituent companies. As described above, continued focus on market risk and volatility is a vicious circle that fosters and creates more correlation and volatility.

Scherer (Sept. 2011) found that the portfolio construction process behind minimum variance investing implicitly picks up risk-based pricing anomalies – ie. the minimum-variance portfolio tends to hold low-beta and low-residual-risk stocks. They also found that 83% of the variation of the minimum variance portfolio excess returns (relative to a capitalization-weighted alternative) can be attributed to the FAMA/FRENCH factors as well as to the returns on two characteristic anomaly portfolios.

**3.5 FTSE/EDHEC Risk adjusted Indices.**

The "Risk-Adjusted" Indices created and launched by FTSE and EDHEC (in December 2009) have inherent problems and are variants of Minimum-variance Index methods. Amenc, Goltz, Martellini & Retkowsky (2010); Martellini (2008); Danielsson, Jorgensen, Sarma & Vries (2006); Danielsson, Jorgensen, Sarma & Vries (2006). Taleb (2008). The results in Daniel and Titman (1997) and Garlappi, Shu & Yan (2008) invalidate the methods and theories in Amenc, Goltz, Martellini & Retkowsky (2010), and show that the FTSE EDHEC Risk-Adjusted indices are not efficient. Indeed the FTSE/EDHEC Risk adjusted Indices have extremely high correlations (typically between 73% and 98%) with related and underlying indices, all of which are market-cap weighted Indices. Chan, Karceski & Lakonishok (1999: 955). This indicates that: a) the emphasis on market risk (standard deviations of returns of stock prices) in the methodology increases calculated correlations, b) the use of a vector that "forces" the sum of all of the Index-weights to be equal to one, distorts the "adjusted risk" and the assigned Index-Weights, c) the "size effect" and "value effects" are dominant and increases correlations among Indices and assets, d) the failure to include operational risk measures as weighting factors increased cross-index correlation and reduced the relevance of the fundamental performance of the constituent companies, e) the "market noise" in stock prices and the returns of stock prices is dominant and causes increased correlations between the risk-adjusted Indices and their underlying market-cap weighted indices. Alexander & Barbosa (2007).

ICAPM/CAPM and Variance/Correlation/Covariance are inaccurate and thus, the FTSE-EDHEC Risk adjusted index methods are not accurate because the "Market portfolio" or the "risk-adjusted" portfolio are not the most efficient portfolio in terms of risk-reward trade-offs; and as shown in this article, the Sharpe Ratio is wrong. Nwogugu (2003); Prono (Jan. 2010); Prono (June 2007); Green & Hollifield (1992); Guo (May/June 2004); Kumar & Ziemba (1993); Lewellen & Nagel (2006). The FTSE EDHEC Risk Efficient Index calculation methods are not accurate because:

a) The "risk-reward tradeoff" is not continuous.

b) The relationship between risk and expected return is not linear as postulated by ICAPM/CAPM, but rather, is non-linear (thus, there may be substantially diminishing increases in returns for every one-unit increase in risk). As explained herein, the *Efficient Frontier* is irrelevant and for portfolios that are above the Efficient Frontier, an increase in risk may not necessarily imply an increase in actual or expected returns. Although the FTSE/EDHEC Risk-Efficient indices are based on the theory of maximizing the Risk-Reward ratio, in reality, the Risk-Reward framework and the mean-variance framework are irrelevant, and are not the primary decision methods of investors and are inefficient. Firstly, some investors don't evaluate opportunities based on risk-reward tradeoffs, and indeed many risk models are based on the absolute allowable or maximum monetary loss. Second, some investors derive greater "transitory" and "final" utilities from low Risk-Reward ratios – this could be attributed to various reasons such as taxes, individual preferences, hedging, currency exposures, etc.. Third, investors are not rational at all times – sentiment, altruism, expectations, habit, biases, prior experiences, personal knowledge, peer influence, herding are all factors that can cause deviation from seeking high risk-reward ratios. Fourth, the risk-reward tradeoff framework does not account for non-monetary utilities gained from holding an asset, selling/buying an asset or having the opportunity to purchase an asset.

**c**) Focusing on variances/Co-variances will always be sub-optimal, because for every "portfolio" that is deemed "Risk-efficient", there is likely to be another portfolio that is more efficient. The typical fund manager does not have constant continuous-time access to the variances/co-variances and expected returns of all public securities (which is a key assumption); and people often disagree about the method for calculating co-variances/variances. Thus, given that investors have different preferences, the "Risk-Efficient" portfolio is not feasible.

**d**) The accuracy of the "Risk-efficient" portfolio will always be inversely proportional to the magnitude of the market noise in the prices and expected returns of securities. ICAPM/CAPM does not capture such market noise efficiently.

**e**) The "Risk-Efficient" portfolio is based on optimization and the Mean-variance framework which has been shown to be inaccurate. In the FTSE/EDHEC Risk-Efficient Index methodology, risk is expressed solely in terms of Standard Deviation (SD) and Variance, whereas the portfolio managers and many risk management systems think and analyze alternatives in terms of the dollar amounts of potential losses (and gains). The method of calculating standard deviations used in the FTSE/EDHEC risk-adjusted Indices, is questionable, and the results are open to various interpretations. Nwogugu (2005a; 2005b). Juttner & Leung (2009). The FTSE EDHEC risk-efficient indices are highly sensitive to errors in Covariances and variances. Kumar & Ziemba (1993).

Alexander & Barbosa (2007) found that on those exchanges where Minimum-Variance hedging may still be more effective than a one-to-one hedge, it is not possible to distinguish which econometric model most efficiently reduces the variance; and that sophisticated econometric models such as GARCH introduce too much noise. The main problem is that Variance ($\sigma^2$) is assumed to be the ideal proxy for risk but there are several limiting factors. Firstly, Variance $\sigma^2$ does not incorporate operational risk or the "fundamental" performance of the company such that the correlation between variance of stock returns on one hand and on the other hand, operational risk and fundamental performance may be negative. Secondly, its conjectured here that for each public company, its total "visible" Variance ($\sigma^2_t$) is the sum of the following:

$\sigma^2_f$ = Variance that is attributable to the fundamental performance and operational risk of the company.

$\sigma^2_m$ = Variance that is attributable to market noise.

$\sigma^2_e$ = Variance that is attributable to market expectations about the company.

$\sigma^2_i$ = Variance that is attributable to industry performance and expectations about the industry.

$\sigma^2_l$ = Variance that is attributable to the volume of trading orders and liquidity of the shares.

$\sigma^2_s$ = Variance that is attributable to the types of investors that own the shares and are interested in buying or selling the shares.

$\sigma^2_o$ = Variance that is attributable to the types of trading orders placed by investors (for purchase of shares of companies in the company's industry) and the trading range of the stock price.

$\sigma^2_c$ = Variance that is attributable to the aggregate changes in the availability of cash for investment by investors, and the cost of margin loans.

$$\sigma^2_f + \sigma^2_m + \sigma^2_e + \sigma^2_i + \sigma^2_l + \sigma^2_s + \sigma^2_o + \sigma^2_c = \sigma^2_t$$

Walsh (1997); Sault (2005); Chan, Karceski & Lakonishok (1999). Unfortunately, the Variance/Standard-Deviation used in the FTSE/EDHEC Risk Adjusted Indices lumps together, all these components of Variance, which is error.

The FTSE/EDHEC "Risk Efficient" Indices cannot reduce or eliminate Index Arbitrage and ETF arbitrage because they don't address these problems. The FTSE EDHEC risk-adjusted indices actually increase Index arbitrage because they constitute tools for market participants to try and exploit perceived un-observed risk premia/discounts inherent in either the original index or the associated Index Futures or the associated FTSE EDHEC Risk-Efficient Index, and to: a) arbitrage among or between the original index and the FTSE-EDHEC Risk-Efficient Index, b) the Index Futures and the FTSE-EDHEC Risk Efficient Index; c) the Index Futures and the original index and the FTSE-EDHEC Risk-Efficient Index. Such arbitrage activities are is essentially gambling because it has no functional purpose, and does not enhance price discovery in markets.

The "efficiency" of the FTSE/EDEHEC Risk-Efficient Indices is measured in terms of the Sharpe Ratio – which is based on standard deviations, and is inaccurate as proved in this document.

The FTSE/EDHEC Risk Efficient Indices increase market volatility in several ways. The typical FTSE EDHEC Risk Efficient Index has a high correlation with the underlying "Base" index. Alexander & Barbosa (2007). The FTSE/EDHEC risk-efficient Indices consider only financial risk but not operational risk; and this creates a vicious circle – the more that market participants rely on financial risk measures (and ignore operational risk and fundamentals), the greater the systemic risk and overall market volatility, and the more the stock market looses its links with the fundamental performance of companies and the greater the market noise. The FTSE/EDHEC risk framework is based on the mean-variance model, which is often totally removed from the actual risk dynamics of companys' daily operations – the Mean-Variance model and the FTSE-EDHEC Indices are "after-the-fact" approaches of toying with historical returns which is a "composite result" (that consist of effects of announced company operating performance, market noise, arbitrage, liquidity, investor preferences, etc.) that does not focus on the operational risk of the subject company. Because they focus almost exclusively on Variances, the Mean-Variance model the FTSE-EDHEC "Risk-Efficient" Indices may be accurate or efficient only if the exact same pattern of historical daily stock prices occur in the future, which is realistically impossible.

The EDHEC Risk-Efficient index erroneously appears to perform better than the underlying Base Index because: a) the FTSE/EDHEC Risk-Efficient index is a sub-set of the Base Index and contains the set of assets that have relatively higher returns and lower variances; b)

**3.6 The Hang Seng Risk-Adjusted Indices.**

The Hang Seng Risk Adjusted Indices ("HSRAI") were developed for investors who are very concerned about volatility. Wong (April 2011). HSRAI are constructed by adjusting the percentages of cash and a "Underlying-Index" (ie. Hang Seng Indices) in a two-asset portfolio such that as volatility increases, the percentage of the portfolio that is cash increases and vice versa according to a specific formula. This embedded adjustment in the formula improperly focuses on the volatility of the Underlying-Index without reference to expectations (about returns) and market noise, which is error. Taleb (2008). Alexander & Barbosa (2007). Haugen and Baker (1991) found that due to investor restrictions on short selling, tax situations, and risk and return expectations, its possible to construct portfolios that dominate the market portfolio in terms of risk-adjusted returns. Thus, the extremely high sensitivity of HSRAI to errors in estimates of Co-variances, variances and realized volatilities renders the HSRAI indices very inaccurate and inappropriate. The HSRAI calculation methods cause a *Volatility Bias* (over-weights cash and securities that have low realized-volatility and under-weights securities that have apparent high volatility). The HSRAI have a *Correlation Bias* – ie. over-weights indices that have a high positive or negative correlation with the "market" or major stock indices; and vice-versa, primarily because the absolute volatility is often compared on a relative basis with market volatility. HSRAI calculation methods have an inherent *Volume Bias* – ie. where Volatility is positively correlated to daily trading volume and or number of completed trades, these Index methods under-weight Indices that are frequently traded and or have high trading volumes; and tend to over-weight Indices that have low daily trading volumes and low number of completed trades). HSRAI index calculation methods have an inherent *Optimization-Rule Bias* – wherein they over-weight or under-weight Indices (in the two-asset portfolio) depending on whether there are weight constraints/caps in the optimization process). HSRAI index calculation methods have an inherent *Substitution Bias* – ie. the probability of substitution of an index (in the two-asset portfolio) for cash is directly proportional to its Correlation to the two-asset portfolio. HSRAI cause a *Noise Indifference Effect* - these methods don't consider the amount of noise in Indices and thus, will under-weight Indices (in the two-asset portfolio) whose prices/values contain significant noise in periods of high-volatility; and will overweight Indices whose prices contain significant noise in periods of low volatility. HSRAI calculation methods cause a *Volatility Estimation Bias* (over-weights or under-weights securities depending on the method used to calculated historical volatility or realized volatility or forecasted volatility. HSRAI calculation methods cause and *Index Calculation Bias* wherein the method of calculation of the Subject Index in the two-asset portfolio (eg. cap-weighted, etc.) affects the HSRAI allocations to cash and calculated values – for example, a Cap-weighted stock Index will tend to overweight low-volatility large cap stocks, and thus, its associated HSRAI index will probably have low cash allocations compared to that of an index that is based on trading-volume.

As mentioned, risk preferences vary among individual investors; and the classic definition of risk as the volatility of total return is inconsistent with investor experience and market trends. As described in Nwogugu (2005a;b), standard deviation is not a good measure of risk. Furthermore, the volatility-based "diversification" relies entirely on market risk, increases correlations and systemic risk, and incorporates substantial market noise and is one more step removed from the fundamental operations and operational risk of the constituent companies in the index. The HSRAI formulas are as follows.

$B_i$ = the level of the Underlying-Index at time i.
$r_t$ = interest rate for the HSRAI for the period t.
TV = Target Volatility set for the HSRAI.
RV = Realized Volatility of the Underlying-Index during the last re-set period.
LF = Hang Seng "Index Exposure" or "Leverage Factor". LF is generally equal to TV/RV. For all periodically re-balanced HSRAI, LF is calculated at each re-balancing date and held constant till the next re-balancing date.
$D_{t,t-1}$ = the difference in days between t and t-1
Cap = maximum "Index Exposure" or "Leverage Factor".
Floor = minimum "Index Exposure" or "Leverage Factor".

$$LF_t = Max\{Min[Cap, (TV/RV_{(t-2)})], Floor\}.$$

$$HSRAI_t = HSRAI_{(t-1)} * \{[1+[LF_t * ((B_t/B_{(t-1)})-1)] + [(1- LF_t) *((r_{(t-1)}/365) * D_{t,(t-1)})]\}$$

Apart from these issues, the HSRAI formulas have the following problems:

**i)** The HSRAI formula does not accurately relate the change in volatility to the change in dynamic allocation between cash and the Underlying-Index. The HSRAI formula erroneously assumes that ∂LF/∂RV is linear, but its really non-linear. Assume TV = 20% and RV = 25%, and thus LF = 80%. If RV= 30%, then LF will be 66.66%. The 5% increase in Volatility resulted in a 13.33% decline in the LF. But it has not been shown and cannot be shown that the 13.33% decline in the LF adequately compensates for the 5% increase in realized volatility – hence the HSRAI has not been calibrated accurately.

Table-1

| TV | RV | ΔRV | LF | ΔLF |
|---|---|---|---|---|
| 20.00% | 8.00% | 0 | 250.0% | 0 |
| 20.00% | 9.00% | 1.00% | 222.2% | -27.778% |
| 20.00% | 10.00% | 1.00% | 200.0% | -22.222% |
| 20.00% | 11.00% | 1.00% | 181.8% | -18.182% |
| 20.00% | 12.00% | 1.00% | 166.7% | -15.152% |
| 20.00% | 13.00% | 1.00% | 153.8% | -12.821% |
| 20.00% | 14.00% | 1.00% | 142.9% | -10.989% |
| 20.00% | 15.00% | 1.00% | 133.3% | -9.524% |
| 20.00% | 16.00% | 1.00% | 125.0% | -8.333% |
| 20.00% | 17.00% | 1.00% | 117.6% | -7.353% |
| 20.00% | 18.00% | 1.00% | 111.1% | -6.536% |
| 20.00% | 19.00% | 1.00% | 105.3% | -5.848% |
| 20.00% | 20.00% | 1.00% | 100.0% | -5.263% |
| 20.00% | 21.00% | 1.00% | 95.2% | -4.762% |
| 20.00% | 22.00% | 1.00% | 90.9% | -4.329% |
| 20.00% | 23.00% | 1.00% | 87.0% | -3.953% |
| 20.00% | 24.00% | 1.00% | 83.3% | -3.623% |
| 20.00% | 25.00% | 1.00% | 80.0% | -3.333% |
| 20.00% | 26.00% | 1.00% | 76.9% | -3.077% |
| 20.00% | 27.00% | 1.00% | 74.1% | -2.849% |
| 20.00% | 28.00% | 1.00% | 71.4% | -2.646% |
| 20.00% | 29.00% | 1.00% | 69.0% | -2.463% |
| 20.00% | 30.00% | 1.00% | 66.7% | -2.299% |
| 20.00% | 31.00% | 1.00% | 64.5% | -2.151% |
| 20.00% | 32.00% | 1.00% | 62.5% | -2.016% |
| 20.00% | 33.00% | 1.00% | 60.6% | -1.894% |
| 20.00% | 34.00% | 1.00% | 58.8% | -1.783% |
| 20.00% | 35.00% | 1.00% | 57.1% | -1.681% |
| 20.00% | 36.00% | 1.00% | 55.6% | -1.587% |
| 20.00% | 37.00% | 1.00% | 54.1% | -1.502% |
| 20.00% | 38.00% | 1.00% | 52.6% | -1.422% |
| 20.00% | 39.00% | 1.00% | 51.3% | -1.350% |
| 20.00% | 40.00% | 1.00% | 50.0% | -1.282% |
| 20.00% | 41.00% | 1.00% | 48.8% | -1.220% |
| 20.00% | 42.00% | 1.00% | 47.6% | -1.161% |
| 20.00% | 43.00% | 1.00% | 46.5% | -1.107% |
| 20.00% | 44.00% | 1.00% | 45.5% | -1.057% |
| 20.00% | 45.00% | 1.00% | 44.4% | -1.010% |
| 20.00% | 46.00% | 1.00% | 43.5% | -0.966% |
| 20.00% | 47.00% | 1.00% | 42.6% | -0.925% |
| 20.00% | 48.00% | 1.00% | 41.7% | -0.887% |

Table-1 clearly shows that the HSRAI is very inefficient because for a 1% increase in Realized Volatility during high-volatility periods (ie. when volatility is 46%-47%), the LF declines by only 0.925%; whereas the same 1% increase in realized volatility during lower-volatility periods (ie. when Realized Volatility is 26%-27%) causes a 2.85% decline in LF. This "***non-monotonic risk sensitivity***" characteristic is a substantial problem; and the opposite should be the case, such that for any 1% increase in Realized Volatility, the LF should decline by much greater amounts during high-volatility periods, than in low-volatility periods.

**ii)** Given that losses or volatility can have utility value, the HSRAI may be suitable only to that universe of investors that don't gain losses from losses or volatility. The HSRAI Index is effectively a new asset class that is significantly removed from the dynamics of the Underlying-Index because of the HSRAI formula – as confirmed by Table-1 above.

**iii)** Cash can be riskier than an Underlying-Index if any of the following conditions exist: **a**) perceived and tax-efficient losses have substantial value to investors; b) the opportunity cost of holding cash is far greater than the opportunity cost of gaining exposure to the Underlying-Index (via exchange traded index options, index futures or direct ownership); **c**) the inflation rate is greater than the yield on cash deposits; **d**) the cash currency depreciates in value against other currencies; **e**) the yield on the cash deposit is fixed; **f)** the cash deposit is not insured and the bank is not stable; **g)** the average utility-value of cash in the economy declines faster than the average utility-value of investments.

**iv)** The HSRAI concept fails if the cost of obtaining exposure to the Underlying-Index (eg. through exchange traded index options and index futures) is less than the cost of the long position in the Underlying-Index. In such circumstances, it will be profitable to simply short the HSRAI index, and then create an equivalent synthetic two-asset portfolio position by buying exchange-traded Index Options/futures and holding cash. This arbitrage activity will tend to increase the volatility of the Underlying-Index, and hence defeats the entire purpose of the HSRAI.

**v**) There is an inherent "*Volatility Leakage Effect*" in the HSRAI Index because the formula for the HSRAI does not include any meaningful or "optimal" time lags between the adjustment dates of the HSRAI Index prices ($D_{t,t-1}$ is the difference in days between t and t-1), so that in periods of extreme volatility, the high volatility in the Underlying-Index will be transmitted to the HSRAI Index will also experience high volatility because after adjustment of the HSRAI index, the absolute value of the two-asset portfolio does not remain constant, and the greater that $D_{t,t-1}$ is, the greater the *Volatility Leakage Effect*. That is, there is a substantial probability that a 1% change in realized volatility causes a greater percentage change in the HSRAI – in Table-1, a 1% increase in realized volatility during low-volatility periods (ie. when Realized volatility is 26%-27%) causes a 2.85% decline in LF, which in turn cause a decline in the expected Value of the HSRAI Index (distinct from current value) that exceeds 1%.

**vi**) When the RV is greater than the TV, then the LF will exceed 100% unless the cap is set at or around the TV. This means that the usefulness of the HSRAI heavily depends on the Target Volatility.

**vii**) In reality, the Target Volatility in the HSRAI model and hence, the purpose of the HSRAI can never be achieved (even when $D_{t,t-1}$ is instantenous) because of the following reasons: **a**) the HSRAI formula is backward-looking and focuses on Realized Volatility, and by the time the HSRAI is revised, the value and volatility of the portion of the HSRAI Index that is the Underlying-Index will have changed; **b**) cash holdings also have inherent volatility primarily from changes in currency exchange rates and the opportunity cost of cash; and thus, the blended volatility of the two-asset portfolio will almost always differ from the Target Volatility (the HSRAI formula erronouusly assumes that cash has no volatility, and that the value of the cash component of the HSRAI will always increase by the applicable interest rate); **c**) the HSRAI formula contains a "weighted averaging" ; **d**) there is an inherent *Representativeness Bias* because although the HSRAI formula is very sensitive to the magnitude of $D_{t,t-1}$, the HSRAI formula does not reflect the importance of $D_{t,t-1}$ especially in the calculation of the term $[1+[LF_t * ((B_t/B_{(t-1)})-1)]$; **e**) if for example $D_{t,t-1}$ is three days and during that period, the Underlying-Index experiences high volatility,

**viii**) Although the HSRAI formula appears to be recursive, its actually non-recursive because: a) $[1+[LF_t * ((Bt_t/B_{(t-1)})-1)]$ does not capture the effect of time on $HSRAI_{(t-1)}$.

**3.7 The S&P Risk Control Index Series (The S&P Developed Market Risk Control Index Series; The S&P Emerging Market Risk Control Indices; And The S&P Global Thematic Risk Control Indices).**

The S&P Risk Control Indices ("SPRCI") were developed for investors who are very concerned about volatility. SPRCI are constructed by adjusting the percentages of cash and a "Underlying-Index" (different Indices) in a two-asset portfolio such that as volatility increases, the percentage of the portfolio that is cash increases and vice versa according to a specific formula. The SPRCI is supposedly dynamically adjusted to "target" a specific level of volatility. Realized historical volatility is calculated using stated methods – eg. an exponentially-weighted average or simple moving average. The SPRCI is typically rebalanced daily. The various SPRCI cover more that fifteen different underlying indices and are summarized in http://www.standardandpoors.com/indices/articles/en/us/?articleType=PDF&assetID=1245206615803; http://www.standardandpoors.com/indices/articles/en/us/?articleType=PDF&assetID=1245198997829. Standard & Poors (August 2011).

This embedded adjustment in the SPRCI formula improperly focuses on the volatility of the Underlying-Index without reference to expectations (about returns) and market noise, which is error. Taleb (2008). Alexander & Barbosa (2007). The SPRCI formula is very similar to the HSRAI formula discussed above, and the two min differences are that: **a)** in the SPRCI formula, the formula for the IE does not include a Floor, while that of the HSRAI includes a floor; and **b)** the formula for calculating the interest on the cash portions of the SPRCI and the HSRAI differs. Thus, the S&P Risk Control Indices have the same weaknesses and problems as the Hang Seng Risk Adjusted Indices which are described above. The SPRCI formula is as follows. Standard & Poors (August 2011: 39-43).

Let:
$\mathbf{B_i}$ = the level of the Underlying-Index at time i.
**rb** = the last Index re-balancing date. The SPRCI's inception date is the first re-balancing date of that SPRCI.
$\mathbf{r_{(i-1)}}$ = stated interest rate for the SPRCI (such as LIBOR or zero).
**TV** = Target Volatility set for the SPRCI.
$\mathbf{RV_{rb}}$= Realized Volatility of the Underlying-Index as of the close of d trading days prior to the previous rebalancing date (rb),where a trading day is defined as a day on which the underlying index is calculated.
$\mathbf{LF_{rb}}$ = "Index exposure" or "Leverage Factor". LF is generally equal to $TV/RV_{(rb-d)}$. For all periodically re-balanced SPRCI, $LF_{rb}$ is calculated at each re-balancing date and held constant till the next re-balancing date.
$\mathbf{D_{t,t-1}}$ = the difference in days between t and t-1. The number of days between the re-balancing date and the date of the volatility reading. eg. if D = 2, then the historical volatility of the Underlying-Index
**Cap** = maximum "Index Exposure" or "Leverage Factor".

$\mathbf{LF_{rb}} = \text{Min}[\text{Cap}, (TV/RV_{(rb-d)})]$

$\mathbf{SPRCI_t} = SPRCI_{(t-1)} *\{1+[LF_{rb} * ((B_t/B_{rb})-1)] + [(1- LF_{rb}) * (\prod^{t}_{(i=rb+1)}\{(1+r_{(i-1)})*(D_{i,(i-1)}/360)\}-1)]\}$

When the SPRCI replicates a rolling investment in a 3-month interest rate, the SPRCI formula is as follows:

$\mathbf{SPRCI_t} = SPRCI_{(t-1)} *\{1+[LF_{rb} * ((B_t/B_{rb})-1)] + [(1- LF_{rb}) * (\prod_{(i=rb+1)}(1+r_{(i-1)})-1)]\}$
Where:
$\mathbf{r_{(i-1)}} = [((D_{i,(i-1)}*IR3M_{(i-1)})* \{(IR3M_{(i-1)} -IR3M_{(i-2)} -D_{(i-1),i}) *( IR3M_{(i-1)} -IR2M_{(i-1)})*(1/30)\}/90)/360]$
$\mathbf{D_{i,(i-1)}}$ = the number of calendar days between day i-1 and day t.
$\mathbf{IR3M_{(i-1)}}$ = the 3-month interest rate on day i-1.
$\mathbf{IR2M_{(i-1)}}$ = the 2-month interest rate on day i-1.

Furthermore, the calculation of Variance (long-term and short-term Realized Volatility) in the SPRCI formula is wrong. Standard & Poors (August 2011: 42-43). The long term volatility and Variance formulas are as follows:

$\text{Long Term Realized Volatility}_{L,t} = \sqrt{\{(252/n)* \text{Variance}_{L,t}\}}$
For **t>T**, $\mathbf{Variance_{L,t}} = \lambda_L* \text{Variance}_{L,t-1} +(1- \lambda_L) *[\ln(B_t/B_{t-n})]^2$
For **t=T**, $\mathbf{Variance_{L,T}} = \Sigma_{i=m+1}(\alpha_{L,i,m}/\text{WeightingFactor}_L) * [\ln(B_i/B_{i-n})]^2$

$\text{Short-term Realized Volatility}_{S,t} = \sqrt{\{(252/n)* \text{Variance}_{S,t}\}}$
For **t>T**, $\mathbf{Variance_{S,t}} = \lambda_S* \text{Variance}_{S,t-1} +(1- \lambda_S) *[\ln(B_t/B_{t-n})]^2$
For **t=T**, $\mathbf{Variance_{S,T}} = \Sigma_{i=m+1} (\alpha_{S,i,m}/\text{WeightingFactor}_S) * [\ln(B_i/B_{i-n})]^2$

Where:
**T** = the start date for the SPRCI.
**n** = number of days inherent in the return calculation used for determining volatility. If n=1, daily returns are used, and if n=2, then two-day returns are used.
**m** = the Nth trading date prior to T.
**N** = the number of trading days observed for calculating the initial variance as of the start date of the index.
$\boldsymbol{\lambda_s}$ = The short term decay factor used for exponential weighting. The decay factor is a number that is greater than zero and less than one; and determines the weight of each daily return in the calculation of historical variance.

$\lambda_L$ = The long term decay factor used for exponential weighting. The decay factor is a number that is greater than zero and less than one; and determines the weight of each daily return in the calculation of historical variance.
**$B_i$ or $B_{i-n}$** = the level of the Underlying Index.
**$\alpha_{L,I,m}$** = weight of date t in the long term volatility calculation, as calculated based on the following formula:

$$\alpha_{L,t} = (1-\lambda_L)^* \lambda_L^{N+m+i}$$

**$\alpha_{S,i,m}$** = weight of date t in the short term volatility calculation, as calculated based on the following formula:

$$\alpha_{S,t} = (1-\lambda_S)^* \lambda_S^{N+m+i}$$

**$WeightingFactor_L$** = $\Sigma^T_{i=m+1}(\alpha_{L,i,m})$
**$WeightingFactor_S$** = $\Sigma^T_{i=m+1}(\alpha_{S,i,m})$

The foregoing Variance formulas are wrong because: **a**) the natural log transformation inherent in the term $\ln(B_t/B_{t-n})$ does not preserve the order of the magnitude of the changes of the periodic returns – the natural log scale differs substantially from most return patterns; **b**) the formula is simply a weighted average of the returns, but does not indicate true variation around a mean or median; **c**) its impractical and inaccurate to assume that the mean of the return series is zero (or that the mean of the natural logs of the return series is zero) – most return series and their natural logs are asymmetrical/skewed around zero; **d**) the weighting factors ($\alpha_{L,t}$, $\alpha_{S,t}$) are biased – because since the decay factors ($\lambda_L$,$\lambda_S$) are between zero and one, as N and m increase, the weighting factors decline, but the opposite should be the case; **e**) weighting the returns on the basis of only time may be useful only if the returns trend/pattern are cyclical or repetitive; and f) time-based weighting may be very mis-leading if there are close periods of very high or low volatility, or if there are frequent intra-period high/low volatility.

The foregoing SPRCI formula erroneously and implicitly assumes that $\partial LF/\partial RV$ is linear, but its really non-linear. Assume TV = 20% and RV = 25%, and thus LF = 80%. If RV= 30%, then LF will be 66.66%. The 5% increase in Volatility resulted in a 13.33% decline in the LF. But it has not been shown and cannot be shown that the 13.33% decline in the LF adequately compensates for the 5% increase in realized volatility – hence the SPRCI formula is not calibrated or accurate. The SPRCI is significantly less efficient than the HSRAI because the SPRCI's Index Exposure (LF) formula does not have any implicit Floor, and thus when TV<RV, there is substantial and disproportionate distortion in the LF - eg. a one percent increase in the RV results in a disproportionately very small decrease in the LF.

## 3.8. The Thomson Reuters Lipper Optimal Target Risk Indices.

Thomson Reuters Lipper Optimal Target Risk Indices (TRLOTRI) are asset allocation-oriented indices designed to assess the trade-off between risk and return in diversified portfolios. The five Target Risk Optimal Indices are Aggressive Growth, Growth, Moderate, Conservative and Very Conservative; and are supposedly are "optimized" based on the Mean-Variance framework and modern portfolio theory. Thomson Reuters (2007).

The Thomson Reuters Lipper Optimal Target Risk Indices are based on the concept that investors try to create "optimal" portfolios in one of two ways: For any level of risk, among a range of available assets, investors consider which of those assets has the same risk and select the one with highest expected return; and for any expected return, among a range of available assets, investors consider which of those assets has the same return and select the one with the lowest risk. These two premises can

Given that the Mean-Variance framework has been shown to be inaccurate, and as explained in proofs above, Variance and Co-Variance are inaccurate measures of risk, the Thomson Reuters Lipper Optimal Target Risk Indices are grossly inaccurate and impractical. Chan, Karceski & Lakonishok (1999). DeMiguel, Garlappi & Uppal (2007); Flam (2010); Prono (June 2009); Taleb (2008); Green & Hollifield (1992). Also, the Mean-Variance framework is highly sensitive to biases in computed returns. Conrad & Kaul (1993).

## 3.9 MSCI Factor Indices.

MSCI created twelve investable indices each of which has exposure to one Barra risk-model factor (eg. factors such as volatility, momentum, etc.) and minimum tracking error relative to the respective underlying standard MSCI Index. The MSCI Risk factor Indices (MSCIRFI) are constructed through optimization, which tries to achieve constant high exposure to a target factor, very low active exposure to all other factors, and minimum tracking error. MSCI (May 2011). To build the MSCI Factor Index, the Parent Index, Benchmark and the relevant Barra Equity Model for the optimization are selected. The optimization is performed from a base currency perspective (e.g., Euro for the MSCI Europe Factor Indices) and allows short selling of securities. For the MSCI Long-Short Factor Indices:

i) the Parent Index is the corresponding MSCI Investable Market Index and serves as the universe of eligible securities for the optimization;
ii) the Benchmark for the optimization is the corresponding MSCI Standard Index; and
iii) the Barra Equity Model is the corresponding global, regional or single country Barra Equity Model.

The MSCIRFI are inaccurate and impractical because: a) as explained in this article, Variance and Covariance are inappropriate and inaccurate measures of risk; b) the limitations/restrictions on the periodic percentage change in the Risk Factor and the weights of the constituent assets in the index eliminates the benefits and purposes of "optimization", and the resulting adverse effects increase as the calculated "weights"; c) the Mean-Variance framework has been shown to be inaccurate and improper; d) stationarity is an implicit assumption of most optimization; but in reality, stationarity does not occur in most instances. Chan, Karceski & Lakonishok (1999). The MSCIRFI index calculation method causes: **i**) a *Volatility Bias* (over-weights securities that have low-volatility and under-weights securities that have high volatility); **ii**) a *Volume Bias* (where Volatility is positively correlated to daily trading volume, this Index method under-weights securities that are frequently traded and over-weights securities that have high daily trading volumes); **iii)** an *Optimization-Rule Bias* (over-weights or under-weights securities depending on whether there are weight constraints/caps in the optimization process); **v)** a *Substitution Bias* (the probability of substitution of a security in the index is directly proportional to its Correlation to the rest of the portfolio); **vi)** a *Noise Indifference Effect* (this index method doesn't consider the amount of noise in security prices; and thus, will under-weight securities whose prices contain significant noise in periods of high-volatility; and will overweight securities whose prices contain significant noise in periods of low volatility; and **v)** a *Volatility Estimation Bias* (over-weights or under-weights securities depending on the method used to calculated historical volatility or realized volatility or forecasted volatility); and **vi)** *Selection/Risk-Reduction Bias* (because the main criteria is the impact of addition of a security on an existing portfolio, the probability of inclusion in the index heavily depends on the nature of already-selected securities).

**3.10 The Dow Jones Relative-Risk Indices.**

Each Dow Jones Relative Risk Index (DJRRI) (Dow Jones Aggressive Portfolio Index; Dow Jones Moderately Aggressive, Dow Jones Moderate, Dow Jones Moderately Conservative and Dow Jones Conservative Portfolio Indices) consists of three Composite Major Asset Classes (CMACs) — stocks, bonds and cash. See: http://www.djindexes.com/mdsidx/downloads/meth_info/Dow_Jones_Relative_Risk_Indexes_Methodology.pdf . The CMACs are represented by specific sub-indices. The stock sub-indexes are products of Dow Jones Indexes; and the bond and cash (T-Bill) sub-indexes are products of Barclays Capital. Within each index, the three CMACs are re-weighted each month to reflect a risk profile that is set at the start of the month based on the current risk level of the stock CMAC. The key elements of the five types of DJRRI are as follows:

**i)** The risk level of the Dow Jones Aggressive Portfolio Index is set monthly to 100% of the current risk of the stock CMAC. The three CMACs are re-weighted within the index to maximize the allocation to the CMAC with the greatest expected return at the 100% risk level.

**ii)** The risk levels of the Dow Jones Moderately Aggressive, Moderate, Moderately Conservative and Conservative Portfolio Indexes are assigned based on the efficient frontier. Once the risk level of the index has been determined each month, the three CMACs are re-weighted within the index to maximize the allocation to the CMAC with the greatest expected return at that risk level.

**iii)** The weighting of each CMAC is not allowed to drop below 5% in any DJRRI.

**iv)** Risk is calculated as 36-month semi-variance. The Percentage of all-stock portfolio risk reflected in the five indices are as follows: Dow Jones Conservative Index – 20%; Dow Jones Moderately Conservative Index – 40%; Dow Jones Moderate Index – 60%; Dow Jones Moderately Aggressive Index – 80%; Dow Jones Aggressive Index – 100%.

For example, the risk level of the Dow Jones Aggressive Index is set monthly to 100% of the current risk of the all-stock portfolio. Risk is calculated as the 36-month semi-variance. The DJRRI are similar to the HSRAI and SPRCI (described above) except that: a) the DJRRI has a three-asset portfolio; a) the assets in the DJRRI portfolio are sub-indices (whereas assets of the HSRAI and SPRCI are cash and a whole-index); c) in the DJRRI "portfolio", the risk of each of the five principal indices is set to a percentage of an "all-stock portfolio".

Hence, the DJRRI are inaccurate and impractical; and has many of the same weaknesses as the HSRAI and SPRCI, some of which are as follows:

**i)** As explained above, Semi-variance and Variance are not accurate or appropriate measures of risk.

**ii)** The efficient frontier and the Mean-variance framework have been shown to be inaccurate. Chan, Karceski & Lakonishok (1999). DeMiguel, Garlappi & Uppal (2007); Flam (2010); Prono (June 2009); Taleb (2008); Green & Hollifield (1992). Also, the Mean-Variance framework is highly sensitive to biases in computed returns. Conrad & Kaul (1993).

***Theorem 7: For Compounded Returns, As The Realized Return Changes From single-digits to Double Digits, the rate of Change of the Standard Deviation increases In A Non-Linear Proportion.***

***Proof***: The proof is a comparison of Table-1 and Table-2 below.

Table-1

| | Time Period 1 | 2 | 3 | 4 | 5 | 6 | 7 | 8 | 9 | 10 | 11 | 12 |
|---|---|---|---|---|---|---|---|---|---|---|---|---|
| | Section-A: | | | | | | | | | | | |
| 1 | 5.0% | 5.0% | 5.0% | 5.0% | 5.0% | 5.0% | 5.0% | 5.0% | 5.0% | 5.0% | 5.0% | 5.0% |
| 2 | 5.0% | 5.0% | 5.0% | 5.0% | 10.0% | 5.0% | 5.0% | 5.0% | 5.0% | 5.0% | 5.0% | 5.0% |
| 3 | 5.0% | 5.0% | 5.0% | 5.0% | 10.0% | 10.0% | 5.0% | 5.0% | 5.0% | 5.0% | 5.0% | 5.0% |
| 4 | 5.0% | 5.0% | 5.0% | 5.0% | 10.0% | 10.0% | 10.0% | 5.0% | 5.0% | 5.0% | 5.0% | 5.0% |
| 5 | 5.0% | 5.0% | 5.0% | 5.0% | 10.0% | 10.0% | 10.0% | 10.0% | 5.0% | 5.0% | 5.0% | 5.0% |
| 6 | 5.0% | 5.0% | 5.0% | 5.0% | 10.0% | 10.0% | 10.0% | 10.0% | 10.0% | 5.0% | 5.0% | 5.0% |
| 7 | 5.0% | 5.0% | 5.0% | 5.0% | 10.0% | 10.0% | 10.0% | 10.0% | 10.0% | 10.0% | 5.0% | 5.0% |
| 8 | 5.0% | 5.0% | 5.0% | 5.0% | 10.0% | 10.0% | 10.0% | 10.0% | 10.0% | 10.0% | 10.0% | 5.0% |
| 9 | 5.0% | 5.0% | 5.0% | 5.0% | 10.0% | 10.0% | 10.0% | 10.0% | 10.0% | 10.0% | 10.0% | 10.0% |
| 10 | 5.0% | 5.0% | 5.0% | 10.0% | 10.0% | 10.0% | 10.0% | 10.0% | 10.0% | 10.0% | 10.0% | 10.0% |
| 11 | 5.0% | 5.0% | 10.0% | 10.0% | 10.0% | 10.0% | 10.0% | 10.0% | 10.0% | 10.0% | 10.0% | 10.0% |
| 12 | 5.0% | 10.0% | 10.0% | 10.0% | 10.0% | 10.0% | 10.0% | 10.0% | 10.0% | 10.0% | 10.0% | 10.0% |
| 13 | 10.0% | 10.0% | 10.0% | 10.0% | 10.0% | 10.0% | 10.0% | 10.0% | 10.0% | 10.0% | 10.0% | 10.0% |
| 14 | -5.0% | -5.0% | -5.0% | -5.0% | -5.0% | -5.0% | -5.0% | -5.0% | -5.0% | -5.0% | -5.0% | -5.0% |
| 15 | -5.0% | -5.0% | -5.0% | -5.0% | -10.0% | -5.0% | -5.0% | -5.0% | -5.0% | -5.0% | -5.0% | -5.0% |
| 16 | -5.0% | -5.0% | -5.0% | -5.0% | -10.0% | -10.0% | -5.0% | -5.0% | -5.0% | -5.0% | -5.0% | -5.0% |
| 17 | -5.0% | -5.0% | -5.0% | -5.0% | -10.0% | -10.0% | -10.0% | -5.0% | -5.0% | -5.0% | -5.0% | -5.0% |
| 18 | -5.0% | -5.0% | -5.0% | -5.0% | -10.0% | -10.0% | -10.0% | -10.0% | -5.0% | -5.0% | -5.0% | -5.0% |
| 19 | -5.0% | -5.0% | -5.0% | -5.0% | -10.0% | -10.0% | -10.0% | -10.0% | -10.0% | -5.0% | -5.0% | -5.0% |
| 20 | -5.0% | -5.0% | -5.0% | -5.0% | -10.0% | -10.0% | -10.0% | -10.0% | -10.0% | -10.0% | -5.0% | -5.0% |
| 21 | -5.0% | -5.0% | -5.0% | -5.0% | -10.0% | -10.0% | -10.0% | -10.0% | -10.0% | -10.0% | -10.0% | -5.0% |
| 22 | -5.0% | -5.0% | -5.0% | -5.0% | -10.0% | -10.0% | -10.0% | -10.0% | -10.0% | -10.0% | -10.0% | -10.0% |
| 23 | -5.0% | -5.0% | -5.0% | -10.0% | -10.0% | -10.0% | -10.0% | -10.0% | -10.0% | -10.0% | -10.0% | -10.0% |
| 24 | -5.0% | -5.0% | -10.0% | -10.0% | -10.0% | -10.0% | -10.0% | -10.0% | -10.0% | -10.0% | -10.0% | -10.0% |
| 25 | -5.0% | -10.0% | -10.0% | -10.0% | -10.0% | -10.0% | -10.0% | -10.0% | -10.0% | -10.0% | -10.0% | -10.0% |
| 26 | -10.0% | -10.0% | -10.0% | -10.0% | -10.0% | -10.0% | -10.0% | -10.0% | -10.0% | -10.0% | -10.0% | -10.0% |
| 27 | 5.0% | 5.0% | 5.0% | 5.0% | 5.0% | 5.0% | 5.0% | 5.0% | 5.0% | 5.0% | 5.0% | 5.0% |
| 28 | 5.0% | 5.0% | 5.0% | 5.0% | -10.0% | 5.0% | 5.0% | 5.0% | 5.0% | 5.0% | 5.0% | 5.0% |
| 29 | 5.0% | 5.0% | 5.0% | 5.0% | -10.0% | -10.0% | 5.0% | 5.0% | 5.0% | 5.0% | 5.0% | 5.0% |
| 30 | 5.0% | 5.0% | 5.0% | 5.0% | -10.0% | -10.0% | -10.0% | 5.0% | 5.0% | 5.0% | 5.0% | 5.0% |
| 31 | 5.0% | 5.0% | 5.0% | 5.0% | -10.0% | -10.0% | -10.0% | -10.0% | 5.0% | 5.0% | 5.0% | 5.0% |
| 32 | 5.0% | 5.0% | 5.0% | 5.0% | -10.0% | -10.0% | -10.0% | -10.0% | -10.0% | 5.0% | 5.0% | 5.0% |
| 33 | 5.0% | 5.0% | 5.0% | 5.0% | -10.0% | -10.0% | -10.0% | -10.0% | -10.0% | -10.0% | 5.0% | 5.0% |
| 34 | 5.0% | 5.0% | 5.0% | 5.0% | -10.0% | -10.0% | -10.0% | -10.0% | -10.0% | -10.0% | -10.0% | 5.0% |
| 35 | 5.0% | 5.0% | 5.0% | 5.0% | -10.0% | -10.0% | -10.0% | -10.0% | -10.0% | -10.0% | -10.0% | -10.0% |
| 36 | 5.0% | 5.0% | 5.0% | -10.0% | -10.0% | -10.0% | -10.0% | -10.0% | -10.0% | -10.0% | -10.0% | -10.0% |
| 37 | 5.0% | 5.0% | -10.0% | -10.0% | -10.0% | -10.0% | -10.0% | -10.0% | -10.0% | -10.0% | -10.0% | -10.0% |
| 38 | 5.0% | -10.0% | -10.0% | -10.0% | -10.0% | -10.0% | -10.0% | -10.0% | -10.0% | -10.0% | -10.0% | -10.0% |
| 39 | -10.0% | -10.0% | -10.0% | -10.0% | -10.0% | -10.0% | -10.0% | -10.0% | -10.0% | -10.0% | -10.0% | -10.0% |

| | Time Period 1 | 2 | 3 | 4 | 5 | 6 | 7 | 8 | |
|---|---|---|---|---|---|---|---|---|---|
| | Section-B: Random Numbers: | | | | | | | | |
| 1 | 5.0% | 3.0% | 9.0% | 6.0% | 1.0% | 3.0% | 7.0% | 9.0% | 2.0 |
| 2 | 2.0% | 8.0% | 2.0% | 9.0% | 40.0% | 3.0% | 2.0% | 5.0% | 4.0 |
| 3 | 4.0% | 1.0% | 9.0% | 6.0% | 10.0% | 11.0% | 4.0% | 1.0% | 7.0 |
| 4 | 5.0% | 3.0% | 5.0% | 7.0% | 36.0% | 15.0% | 10.0% | 8.0% | 5.0 |
| 5 | 8.0% | 1.0% | 9.0% | 6.0% | 27.0% | 40.0% | 12.0% | 13.0% | 9.0 |
| 6 | 5.0% | 6.0% | 3.0% | 3.0% | 19.0% | 11.0% | 19.0% | 10.0% | 10.0 |
| 7 | 9.0% | 3.0% | 9.0% | 6.0% | 10.0% | 24.0% | 10.0% | 34.0% | 22.0 |
| 8 | 5.0% | 1.0% | 6.0% | 1.0% | 12.0% | 10.0% | 35.0% | 10.0% | 11.0 |
| 9 | 1.0% | 3.0% | 9.0% | 6.0% | 10.0% | 32.0% | 14.0% | 65.0% | 19.0 |
| 10 | 5.0% | 3.0% | 2.0% | 13.0% | 34.0% | 11.0% | 23.0% | 15.0% | 41.0 |
| 11 | 6.0% | 3.0% | 15.0% | 46.0% | 53.0% | 56.0% | 78.0% | 10.0% | 13.0 |
| 12 | 5.0% | 22.0% | 11.0% | 22.0% | 10.0% | 23.0% | 11.0% | 11.0% | 10.0 |
| 13 | 10.0% | 17.0% | 10.0% | 13.0% | 16.0% | 11.0% | 17.0% | 31.0% | 12.0 |
| | Random Numbers: | | | | | | | | |
| 14 | -5.0% | -9.0% | -2.0% | -6.0% | -5.0% | -1.0% | -4.0% | -3.0% | -8.0 |
| 15 | -2.0% | -9.0% | -1.0% | -4.0% | -10.0% | -1.0% | -1.0% | -7.0% | -2.0 |
| 16 | -5.0% | -1.0% | -8.0% | -6.0% | -11.0% | -32.0% | -4.0% | -3.0% | -8.0 |
| 17 | -6.0% | -9.0% | -2.0% | -9.0% | -10.0% | -10.0% | -23.0% | -9.0% | -4.0 |
| 18 | -5.0% | -8.0% | -4.0% | -6.0% | -34.0% | -13.0% | -10.0% | -56.0% | -8.0 |
| 19 | -9.0% | -9.0% | -2.0% | -8.0% | -11.0% | -10.0% | -15.0% | -10.0% | 63.0 |
| 20 | -5.0% | -3.0% | -6.0% | -6.0% | -10.0% | -15.0% | -10.0% | -14.0% | 10.0 |
| 21 | -1.0% | -9.0% | -2.0% | -1.0% | -44.0% | -10.0% | -10.0% | -10.0% | 20.0 |
| 22 | -5.0% | -6.0% | -1.0% | -6.0% | -10.0% | -65.0% | -16.0% | -77.0% | 14.0 |
| 23 | -3.0% | -9.0% | -2.0% | -10.0% | -33.0% | -10.0% | -10.0% | -11.0% | 10.0 |
| 24 | -5.0% | -5.0% | -10.0% | -22.0% | -10.0% | -34.0% | -33.0% | -10.0% | 11.0 |
| 25 | -4.0% | -10.0% | -14.0% | -10.0% | -13.0% | -22.0% | -10.0% | -16.0% | 10.0 |
| 26 | -33.0% | -11.0% | -35.0% | -12.0% | -10.0% | -10.0% | -15.0% | -10.0% | 22.0 |
| | Random Numbers: | | | | | | | | |
| 27 | 5.0% | 7.0% | 4.0% | 3.0% | 9.0% | 2.0% | 5.0% | 8.0% | 7.0 |
| 28 | 2.0% | 5.0% | 1.0% | 1.0% | -10.0% | 2.0% | 5.0% | 1.0% | 11.0 |
| 29 | 5.0% | 7.0% | 4.0% | 9.0% | -11.0% | -19.0% | 5.0% | 6.0% | 2.0 |
| 30 | 6.0% | 4.0% | 9.0% | 3.0% | -16.0% | -14.0% | -34.0% | 8.0% | 5.0 |
| 31 | 5.0% | 7.0% | 4.0% | 1.0% | -19.0% | -20.0% | -11.0% | -16.0% | 7.0 |
| 32 | 8.0% | 2.0% | 5.0% | 3.0% | -13.0% | -65.0% | -19.0% | -22.0% | 44.0 |
| 33 | 5.0% | 7.0% | 4.0% | 5.0% | -10.0% | -17.0% | -33.0% | -18.0% | 10.0 |
| 34 | 1.0% | 8.0% | 3.0% | 3.0% | -11.0% | -61.0% | -19.0% | -10.0% | 27.0 |
| 35 | 5.0% | 7.0% | 4.0% | 7.0% | -13.0% | -10.0% | -20.0% | -10.0% | 18.0 |
| 36 | 9.0% | 3.0% | 8.0% | -13.0% | -10.0% | -19.0% | -10.0% | -11.0% | 10.0 |
| 37 | 5.0% | 7.0% | -20.0% | -14.0% | -18.0% | -10.0% | -15.0% | -10.0% | 56.0 |
| 38 | 2.0% | -10.0% | -11.0% | -53.0% | -10.0% | -16.0% | -10.0% | -22.0% | 10.0 |
| 39 | -10.0% | -10.0% | -10.0% | -10.0% | -13.0% | -10.0% | -10.0% | -10.0% | 10.0 |

| | Section-A Returns | First Cross-Sectional Difference {(R(n+1) - R(n)} | Section-A Standard Deviation | First Cross-Sectional Difference {(σ(n+1) - σ(n)} |
|---|---|---|---|---|
| R1 | 79.59% | 0.00% | 0.000% | 0.00% |
| R2 | 88.14% | -8.55% | 1.443% | -1.44% |
| R3 | 97.10% | -8.96% | 1.946% | -0.50% |
| R4 | 106.48% | -9.39% | 2.261% | -0.32% |
| R5 | 116.31% | -9.83% | 2.462% | -0.20% |
| R6 | 126.61% | -10.30% | 2.575% | -0.11% |
| R7 | 137.41% | -10.79% | 2.611% | -0.04% |
| R8 | 148.71% | -11.31% | 2.575% | 0.04% |
| R9 | 160.55% | -11.84% | 2.462% | 0.11% |
| R10 | 172.96% | -12.41% | 2.261% | 0.20% |
| R11 | 185.96% | -13.00% | 1.946% | 0.32% |
| R12 | 199.58% | -13.62% | 1.443% | 0.50% |
| R13 | 213.84% | -14.27% | 0.000% | 1.44% |
| R14 | -45.96% | 0.00% | 0.000% | 0.00% |
| R15 | -48.81% | 2.84% | 1.443% | -1.44% |
| R16 | -51.50% | 2.69% | 1.946% | -0.50% |
| R17 | -54.05% | 2.55% | 2.261% | -0.32% |
| R18 | -56.47% | 2.42% | 2.462% | -0.20% |
| R19 | -58.76% | 2.29% | 2.575% | -0.11% |
| R20 | -60.93% | 2.17% | 2.611% | -0.04% |
| R21 | -62.99% | 2.06% | 2.575% | 0.04% |
| R22 | -64.94% | 1.95% | 2.462% | 0.11% |
| R23 | -66.78% | 1.85% | 2.261% | 0.20% |
| R24 | -68.53% | 1.75% | 1.946% | 0.32% |
| R25 | -70.19% | 1.66% | 1.443% | 0.50% |
| R26 | -71.76% | 1.57% | 0.000% | 1.44% |
| R27 | 79.59% | 0.00% | 0.000% | 0.00% |
| R28 | 53.93% | 25.66% | 4.330% | -4.33% |
| R29 | 31.94% | 21.99% | 5.839% | -1.51% |

| | Section-B (Random Numbers) Returns | First Cross-Sectional Difference {(R(n+1) - R(n)} | Section-B (Random Numbers): Standard Deviation | First Cross-Sectional Difference {(σ(n+1) - σ(n)} |
|---|---|---|---|---|
| R1 | 79.59% | 0.00% | 2.807% | 0.00% |
| R2 | 88.14% | -8.55% | 10.760% | -7.95% |
| R3 | 97.10% | -8.96% | 3.384% | 7.38% |
| R4 | 106.48% | -9.39% | 9.064% | -5.68% |
| R5 | 116.31% | -9.83% | 11.229% | -2.17% |
| R6 | 126.61% | -10.30% | 5.954% | 5.27% |
| R7 | 137.41% | -10.79% | 9.139% | -3.18% |
| R8 | 148.71% | -11.31% | 10.603% | -1.46% |
| R9 | 160.55% | -11.84% | 17.270% | -6.67% |
| R10 | 172.96% | -12.41% | 11.935% | 5.34% |
| R11 | 185.96% | -13.00% | 24.766% | -12.83% |
| R12 | 199.58% | -13.62% | 7.180% | 17.59% |
| R13 | 213.84% | -14.27% | 6.097% | 1.08% |
| R14 | -45.96% | 0.00% | 2.730% | 0.00% |
| R15 | -48.81% | 2.84% | 3.397% | -0.67% |
| R16 | -51.50% | 2.69% | 7.987% | -4.59% |
| R17 | -54.05% | 2.55% | 5.726% | 2.26% |
| R18 | -56.47% | 2.42% | 15.710% | -9.98% |
| R19 | -58.76% | 2.29% | 16.261% | -0.55% |
| R20 | -60.93% | 2.17% | 20.273% | -4.01% |
| R21 | -62.99% | 2.06% | 13.667% | 6.61% |
| R22 | -64.94% | 1.95% | 24.385% | -10.72% |
| R23 | -66.78% | 1.85% | 7.982% | 16.40% |
| R24 | -68.53% | 1.75% | 9.893% | -1.91% |
| R25 | -70.19% | 1.66% | 7.489% | 2.40% |
| R26 | -71.76% | 1.57% | 10.260% | -2.77% |
| R27 | 79.59% | 0.00% | 2.179% | 0.00% |
| R28 | 53.93% | 25.66% | 4.887% | -2.71% |
| R29 | 31.94% | 21.99% | 8.284% | -3.40% |
| R30 | 13.09% | 18.85% | 6.784% | -0.95% |
| R31 | -3.06% | 16.16% | 7.385% | -0.60% |
| R32 | -16.91% | 13.85% | 7.724% | -0.34% |
| R33 | -28.78% | 11.87% | 7.833% | -0.11% |
| R34 | -38.96% | 10.17% | 7.724% | 0.11% |
| R35 | -47.68% | 8.72% | 7.385% | 0.34% |
| R36 | -55.15% | 7.47% | 6.784% | 0.60% |
| R37 | -61.56% | 6.41% | 5.839% | 0.95% |
| R38 | -67.05% | 5.49% | 4.330% | 1.51% |
| R39 | -71.76% | 4.71% | 0.000% | 4.33% |
| R30 | 13.09% | 18.85% | 13.266% | -4.98% |
| R31 | -3.06% | 16.16% | 10.673% | 2.59% |
| R32 | -16.91% | 13.85% | 23.240% | -12.57% |
| R33 | -28.78% | 11.87% | 12.831% | 10.41% |
| R34 | -38.96% | 10.17% | 19.205% | -6.37% |
| R35 | -47.68% | 8.72% | 14.727% | 4.48% |
| R36 | -55.15% | 7.47% | 10.235% | 4.49% |
| R37 | -61.56% | 6.41% | 16.714% | -6.48% |
| R38 | -67.05% | 5.49% | 13.263% | 3.45% |
| R39 | -71.76% | 4.71% | 0.866% | 12.40% |

Table-2

| | Time Period | | | | | | | | | | | |
|---|---|---|---|---|---|---|---|---|---|---|---|---|
| | 1 | 2 | 3 | 4 | 5 | 6 | 7 | 8 | 9 | 10 | 11 | 12 |
| | Section-A: | | | | | | | | | | | |
| 1 | 110.0% | 110.0% | 110.0% | 110.0% | 110.0% | 110.0% | 110.0% | 110.0% | 110.0% | 110.0% | 110.0% | 110.0% |
| 2 | 110.0% | 110.0% | 110.0% | 110.0% | 10.0% | 110.0% | 110.0% | 110.0% | 110.0% | 110.0% | 110.0% | 110.0% |
| 3 | 110.0% | 110.0% | 110.0% | 110.0% | 10.0% | 10.0% | 110.0% | 110.0% | 110.0% | 110.0% | 110.0% | 110.0% |
| 4 | 110.0% | 110.0% | 110.0% | 110.0% | 10.0% | 10.0% | 10.0% | 110.0% | 110.0% | 110.0% | 110.0% | 110.0% |
| 5 | 110.0% | 110.0% | 110.0% | 110.0% | 10.0% | 10.0% | 10.0% | 10.0% | 110.0% | 110.0% | 110.0% | 110.0% |
| 6 | 110.0% | 110.0% | 110.0% | 110.0% | 10.0% | 10.0% | 10.0% | 10.0% | 10.0% | 110.0% | 110.0% | 110.0% |
| 7 | 110.0% | 110.0% | 110.0% | 110.0% | 10.0% | 10.0% | 10.0% | 10.0% | 10.0% | 10.0% | 110.0% | 110.0% |
| 8 | 110.0% | 110.0% | 110.0% | 110.0% | 10.0% | 10.0% | 10.0% | 10.0% | 10.0% | 10.0% | 10.0% | 110.0% |
| 9 | 110.0% | 110.0% | 110.0% | 110.0% | 10.0% | 10.0% | 10.0% | 10.0% | 10.0% | 10.0% | 10.0% | 10.0% |
| 10 | 110.0% | 110.0% | 110.0% | 10.0% | 10.0% | 10.0% | 10.0% | 10.0% | 10.0% | 10.0% | 10.0% | 10.0% |
| 11 | 110.0% | 110.0% | 10.0% | 10.0% | 10.0% | 10.0% | 10.0% | 10.0% | 10.0% | 10.0% | 10.0% | 10.0% |
| 12 | 110.0% | 10.0% | 10.0% | 10.0% | 10.0% | 10.0% | 10.0% | 10.0% | 10.0% | 10.0% | 10.0% | 10.0% |
| 13 | 10.0% | 10.0% | 10.0% | 10.0% | 10.0% | 10.0% | 10.0% | 10.0% | 10.0% | 10.0% | 10.0% | 10.0% |
| 14 | -110.0% | -110.0% | -110.0% | -110.0% | -110.0% | -110.0% | -110.0% | -110.0% | -110.0% | -110.0% | -110.0% | -110.0% |
| 15 | -110.0% | -110.0% | -110.0% | -110.0% | 10.0% | -110.0% | -110.0% | -110.0% | -110.0% | -110.0% | -110.0% | -110.0% |
| 16 | -110.0% | -110.0% | -110.0% | -110.0% | 10.0% | 10.0% | -110.0% | -110.0% | -110.0% | -110.0% | -110.0% | -110.0% |
| 17 | -110.0% | -110.0% | -110.0% | -110.0% | 10.0% | 10.0% | 10.0% | -110.0% | -110.0% | -110.0% | -110.0% | -110.0% |
| 18 | -110.0% | -110.0% | -110.0% | -110.0% | 10.0% | 10.0% | 10.0% | 10.0% | -110.0% | -110.0% | -110.0% | -110.0% |
| 19 | -110.0% | -110.0% | -110.0% | -110.0% | 10.0% | 10.0% | 10.0% | 10.0% | 10.0% | -110.0% | -110.0% | -110.0% |
| 20 | -110.0% | -110.0% | -110.0% | -110.0% | 10.0% | 10.0% | 10.0% | 10.0% | 10.0% | 10.0% | -110.0% | -110.0% |
| 21 | -110.0% | -110.0% | -110.0% | -110.0% | 10.0% | 10.0% | 10.0% | 10.0% | 10.0% | 10.0% | 10.0% | -110.0% |
| 22 | -110.0% | -110.0% | -110.0% | -110.0% | 10.0% | 10.0% | 10.0% | 10.0% | 10.0% | 10.0% | 10.0% | 10.0% |
| 23 | -110.0% | -110.0% | -110.0% | 10.0% | 10.0% | 10.0% | 10.0% | 10.0% | 10.0% | 10.0% | 10.0% | 10.0% |
| 24 | -110.0% | -110.0% | 10.0% | 10.0% | 10.0% | 10.0% | 10.0% | 10.0% | 10.0% | 10.0% | 10.0% | 10.0% |
| 25 | -110.0% | 10.0% | 10.0% | 10.0% | 10.0% | 10.0% | 10.0% | 10.0% | 10.0% | 10.0% | 10.0% | 10.0% |
| 26 | 10.0% | 10.0% | 10.0% | 10.0% | 10.0% | 10.0% | 10.0% | 10.0% | 10.0% | 10.0% | 10.0% | 10.0% |
| 27 | -110.0% | -110.0% | -110.0% | -110.0% | -110.0% | -110.0% | -110.0% | -110.0% | -110.0% | -110.0% | -110.0% | -110.0% |
| 28 | -110.0% | -110.0% | -110.0% | -110.0% | -10.0% | -110.0% | -110.0% | -110.0% | -110.0% | -110.0% | -110.0% | -110.0% |
| 29 | -110.0% | -110.0% | -110.0% | -110.0% | -10.0% | -10.0% | -110.0% | -110.0% | -110.0% | -110.0% | -110.0% | -110.0% |
| 30 | -110.0% | -110.0% | -110.0% | -110.0% | -10.0% | -10.0% | -10.0% | -110.0% | -110.0% | -110.0% | -110.0% | -110.0% |
| 31 | -110.0% | -110.0% | -110.0% | -110.0% | -10.0% | -10.0% | -10.0% | -10.0% | -110.0% | -110.0% | -110.0% | -110.0% |
| 32 | -110.0% | -110.0% | -110.0% | -110.0% | -10.0% | -10.0% | -10.0% | -10.0% | -10.0% | -110.0% | -110.0% | -110.0% |
| 33 | -110.0% | -110.0% | -110.0% | -110.0% | -10.0% | -10.0% | -10.0% | -10.0% | -10.0% | -10.0% | -110.0% | -110.0% |
| 34 | -110.0% | -110.0% | -110.0% | -110.0% | -10.0% | -10.0% | -10.0% | -10.0% | -10.0% | -10.0% | -10.0% | -110.0% |
| 35 | -110.0% | -110.0% | -110.0% | -110.0% | -10.0% | -10.0% | -10.0% | -10.0% | -10.0% | -10.0% | -10.0% | -10.0% |
| 36 | -110.0% | -110.0% | -110.0% | -10.0% | -10.0% | -10.0% | -10.0% | -10.0% | -10.0% | -10.0% | -10.0% | -10.0% |
| 37 | -110.0% | -110.0% | -10.0% | -10.0% | -10.0% | -10.0% | -10.0% | -10.0% | -10.0% | -10.0% | -10.0% | -10.0% |
| 38 | -110.0% | -10.0% | -10.0% | -10.0% | -10.0% | -10.0% | -10.0% | -10.0% | -10.0% | -10.0% | -10.0% | -10.0% |
| 39 | -10.0% | -10.0% | -10.0% | -10.0% | -10.0% | -10.0% | -10.0% | -10.0% | -10.0% | -10.0% | -10.0% | -10.0% |

| | Time Period | | | | | | |
|---|---|---|---|---|---|---|---|
| | 1 | 2 | 3 | 4 | 5 | 6 | 7 |
| | Section-B: Random Numbers: | | | | | | |
| 1 | 110.0% | 121.0% | 101.0% | 133.0% | 154.0% | 171.0% | 119.0% |
| 2 | 121.0% | 189.0% | 185.0% | 110.0% | 10.0% | 135.0% | 211.0% |
| 3 | 110.0% | 111.0% | 122.0% | 171.0% | 11.0% | 35.0% | 128.0% |
| 4 | 156.0% | 121.0% | 101.0% | 133.0% | 34.0% | 12.0% | 61.0% |
| 5 | 110.0% | 231.0% | 132.0% | 222.0% | 28.0% | 10.0% | 15.0% |
| 6 | 298.0% | 121.0% | 101.0% | 133.0% | 14.0% | 34.0% | 56.0% |
| 7 | 110.0% | 322.0% | 151.0% | 237.0% | 65.0% | 10.0% | 11.0% |
| 8 | 132.0% | 121.0% | 101.0% | 133.0% | 16.0% | 22.0% | 40.0% |
| 9 | 110.0% | 117.0% | 333.0% | 191.0% | 10.0% | 15.0% | 22.0% |
| 10 | 221.0% | 121.0% | 101.0% | 11.0% | 25.0% | 32.0% | 45.0% |
| 11 | 110.0% | 112.0% | 10.0% | 32.0% | 16.0% | 54.0% | 22.0% |
| 12 | 175.0% | 10.0% | 11.0% | 18.0% | 30.0% | 52.0% | 41.0% |
| 13 | 10.0% | 12.0% | 36.0% | 60.0% | 15.0% | 25.0% | 40.0% |
| | Random Numbers: | | | | | | |
| 14 | -110.0% | -121.0% | -101.0% | -133.0% | -154.0% | -171.0% | -119.0% |
| 15 | -121.0% | -189.0% | -185.0% | -110.0% | 10.0% | -135.0% | -211.0% |
| 16 | -110.0% | -111.0% | -122.0% | -171.0% | 11.0% | 35.0% | -128.0% |
| 17 | -156.0% | -121.0% | -101.0% | -133.0% | 34.0% | 12.0% | 61.0% |
| 18 | -110.0% | -231.0% | -132.0% | -222.0% | 28.0% | 10.0% | 15.0% |
| 19 | -298.0% | -121.0% | -101.0% | -133.0% | 14.0% | 34.0% | 56.0% |
| 20 | -110.0% | -322.0% | -151.0% | -237.0% | 65.0% | 10.0% | 11.0% |
| 21 | -132.0% | -121.0% | -101.0% | -133.0% | 16.0% | 22.0% | 40.0% |
| 22 | -110.0% | -117.0% | -333.0% | -191.0% | 10.0% | 15.0% | 22.0% |
| 23 | -221.0% | -121.0% | -101.0% | 11.0% | 25.0% | 32.0% | 45.0% |
| 24 | -110.0% | -112.0% | 10.0% | 32.0% | 16.0% | 54.0% | 22.0% |
| 25 | -175.0% | 10.0% | 11.0% | 18.0% | 30.0% | 52.0% | 41.0% |
| 26 | 10.0% | 12.0% | 36.0% | 60.0% | 15.0% | 25.0% | 40.0% |
| | Random Numbers: | | | | | | |
| 27 | -110.0% | -121.0% | -101.0% | -133.0% | -154.0% | -171.0% | -119.0% |
| 28 | -121.0% | -189.0% | -185.0% | -110.0% | -10.0% | -171.0% | -133.0% |
| 29 | -110.0% | -111.0% | -122.0% | -171.0% | -11.0% | -35.0% | -151.0% |
| 30 | -156.0% | -121.0% | -101.0% | -133.0% | -34.0% | -12.0% | -61.0% |
| 31 | -110.0% | -231.0% | -132.0% | -222.0% | -28.0% | -10.0% | -15.0% |
| 32 | -298.0% | -121.0% | -101.0% | -133.0% | -14.0% | -34.0% | -56.0% |
| 33 | -110.0% | -322.0% | -151.0% | -237.0% | -65.0% | -10.0% | -11.0% |
| 34 | -132.0% | -121.0% | -101.0% | -133.0% | -16.0% | -22.0% | -40.0% |
| 35 | -110.0% | -117.0% | -333.0% | -191.0% | -10.0% | -15.0% | -22.0% |
| 36 | -221.0% | -121.0% | -101.0% | -11.0% | -25.0% | -32.0% | -45.0% |
| 37 | -110.0% | -112.0% | -10.0% | -32.0% | -16.0% | -54.0% | -22.0% |
| 38 | -175.0% | -10.0% | -11.0% | -18.0% | -30.0% | -52.0% | -41.0% |
| 39 | -10.0% | -12.0% | -36.0% | -60.0% | -15.0% | -25.0% | -40.0% |

| | Section-A Returns | First Cross-Sectional Difference {(R(n+1) - R(n)} | Section-A Standard Deviation | First Cross-Sectional Difference {(σ(n+1) - σ(n)} |
|---|---|---|---|---|
| R1 | 735482.8% | 0.0% | 0.00% | 0.00% |
| R2 | 385205.3% | 350277.5% | 28.87% | -28.87% |
| R3 | 201726.6% | 183478.7% | 38.92% | -10.06% |
| R4 | 105618.7% | 96107.9% | 45.23% | -6.30% |
| R5 | 55276.4% | 50342.2% | 49.24% | -4.01% |
| R6 | 28906.7% | 26369.7% | 51.49% | -2.26% |
| R7 | 15094.0% | 13812.7% | 52.22% | -0.73% |
| R8 | 7858.8% | 7235.2% | 51.49% | 0.73% |
| R9 | 4068.9% | 3789.9% | 49.24% | 2.26% |
| R10 | 2083.7% | 1985.2% | 45.23% | 4.01% |
| R11 | 1043.8% | 1039.9% | 38.92% | 6.30% |
| R12 | 499.2% | 544.7% | 28.87% | 10.06% |
| R13 | 213.8% | 285.3% | 0.00% | 28.87% |
| R14 | -100.000% | 0.000% | 0.00% | 0.00% |
| R15 | -100.000% | 0.000% | 34.64% | -34.64% |
| R16 | -100.000% | 0.000% | 46.71% | -12.07% |
| R17 | -100.000% | 0.000% | 54.27% | -7.56% |
| R18 | -100.000% | 0.000% | 59.08% | -4.81% |
| R19 | -100.000% | 0.000% | 61.79% | -2.71% |
| R20 | -100.000% | 0.000% | 62.67% | -0.88% |
| R21 | -100.002% | 0.002% | 61.79% | 0.88% |
| R22 | -99.979% | -0.023% | 59.08% | 2.71% |
| R23 | -100.236% | 0.257% | 54.27% | 4.81% |
| R24 | -97.406% | -2.830% | 46.71% | 7.56% |
| R25 | -128.531% | 31.125% | 34.64% | 12.07% |
| R26 | 213.843% | -342.374% | 0.00% | 34.64% |
| R27 | -100.000% | 0.000% | 0.00% | 0.00% |

| | Section-B (Random Numbers) Returns | First Cross-Sectional Difference {(R(n+1) - R(n)} | Section-B (Random Numbers): Standard Deviation | First Cross-Sectional Difference {(σ(n+1) - σ(n)} |
|---|---|---|---|---|
| R1 | 2927104.78% | 0.00% | 27.930% | 0.00% |
| R2 | 5847495.29% | -2920390.51% | 99.652% | -71.72% |
| R3 | 1316465.01% | 4531030.28% | 60.882% | 38.77% |
| R4 | 807640.79% | 508824.22% | 83.715% | -22.83% |
| R5 | 400435.70% | 407205.09% | 80.148% | 3.57% |
| R6 | 222098.20% | 178337.50% | 82.767% | -2.62% |
| R7 | 149635.50% | 72462.71% | 98.704% | -15.94% |
| R8 | 42033.49% | 107602.00% | 58.316% | 40.39% |
| R9 | 41973.66% | 59.83% | 96.315% | -38.00% |
| R10 | 10415.39% | 31558.27% | 62.409% | 33.91% |
| R11 | 4345.26% | 6070.14% | 35.390% | 27.02% |
| R12 | 3188.62% | 1156.64% | 46.038% | -10.65% |
| R13 | 1964.16% | 1224.46% | 19.256% | 26.78% |
| R14 | -100.000% | 0.000% | 27.930% | 0.00% |
| R15 | -100.005% | 0.005% | 102.609% | -74.68% |
| R16 | -99.993% | -0.012% | 76.187% | 26.42% |
| R17 | -100.000% | 0.008% | 107.218% | -31.03% |
| R18 | -99.550% | -0.451% | 101.716% | 5.50% |
| R19 | -100.013% | 0.464% | 109.454% | -7.74% |
| R20 | -95.083% | -4.930% | 122.081% | -12.63% |
| R21 | -100.115% | 5.032% | 88.392% | 33.69% |
| R22 | -73.588% | -26.527% | 121.120% | -32.73% |
| R23 | -101.874% | 28.286% | 83.320% | 37.80% |
| R24 | -88.018% | -13.856% | 55.153% | 28.17% |
| R25 | -996.896% | 908.878% | 60.476% | -5.32% |
| R26 | 1964.156% | -2961.052% | 19.256% | 41.22% |
| R27 | -100.000% | 0.000% | 27.930% | 0.00% |

| | | | | | | | | | |
|---|---|---|---|---|---|---|---|---|---|
| R28 | -100.000% | 0.000% | 28.87% | -28.87% | R28 | -100.004% | 0.004% | 79.121% | -51.19% |
| R29 | -100.000% | 0.000% | 38.92% | -10.06% | R29 | -99.957% | -0.046% | 66.500% | 12.62% |
| R30 | -100.000% | 0.000% | 45.23% | -6.30% | R30 | -100.011% | 0.054% | 80.940% | -14.44% |
| R31 | -100.000% | 0.000% | 49.24% | -4.01% | R31 | -99.978% | -0.033% | 75.042% | 5.90% |
| R32 | -100.000% | 0.000% | 51.49% | -2.26% | R32 | -100.000% | 0.022% | 77.692% | -2.65% |
| R33 | -100.000% | 0.000% | 52.22% | -0.73% | R33 | -98.524% | -1.476% | 107.043% | -29.35% |
| R34 | -100.000% | 0.001% | 51.49% | 0.73% | R34 | -100.001% | 1.477% | 59.538% | 47.50% |
| R35 | -99.996% | -0.005% | 49.24% | 2.26% | R35 | -99.875% | -0.126% | 96.315% | -36.78% |
| R36 | -100.039% | 0.043% | 45.23% | 4.01% | R36 | -100.017% | 0.142% | 62.409% | 33.91% |
| R37 | -99.651% | -0.387% | 38.92% | 6.30% | R37 | -99.955% | -0.062% | 35.390% | 27.02% |
| R38 | -103.138% | 3.487% | 28.87% | 10.06% | R38 | -101.872% | 1.917% | 46.038% | -10.65% |
| R39 | -71.757% | -31.381% | 0.00% | 28.87% | R39 | -99.189% | -2.683% | 19.256% | 26.78% |

***Theorem 8: For Compounded Returns, As The Realized Returns Change From Double-digits to Triple-Digits, the rate of Change of the Standard Deviation increases in non-linear proportions and exponentially.***
***Proof***: The proof is evident by comparing the compounded returns in Table-2 above.□

***Theorem-9: All Of The Above-Mentioned Biases Are "Matching" Biases.***
*Proof*: Amenc, Goltz, Martellini & Retkowsky (2010), Amenc, Goltz & Le Sourd (Sept. 2006), Choueifaty & Coignard (2008), Clarke, de Silva & Thorley (2006), Fernholz, Garvy & Hannon (1998), Maillard, Roncalli & Teiletche (2008), Martellini (2008), Qian (Sept. 2005), and Qian (2006) describe the formulas for calculating the above mentioned indices. Let I be the set of all possible index values, and $I_t$ be the Index value at time t. B is the set of all values of a Bias, and $B_t$ is the value of the B at time *t*. $B_t$ can take on positive or negative values. Then for any time interval *t+n*, and in any market condition, given the formulas for the Indices, any change $\Delta I$ causes an automatic change ($\Delta B$) in B. Hence, these Biases are "matching" Biases because any change in the Index value $I_t$ creates a change in the Bias $B_t$.□

***Theorem-10: The Above-Mentioned Biases Are Recursive.***
*Proof*: Amenc, Goltz & Le Sourd (Sept. 2006), Choueifaty & Coignard (2008), Clarke, de Silva & Thorley (2006), Fernholz, Garvy & Hannon (1998), Maillard, Roncalli & Teiletche (2008), Martellini (2008), Qian (Sept. 2005), and Qian (2006) describe the formulas for calculating the above mentioned indices. Backus, Routledge & Zin (December 2005), and Epaulard & Pommeret (Jan. 2001) Klibanoff P, Marinacci M & Mukerji S (2009) discuss recursive preferences. Let $I_i$ be the set of all possible index values, and $I_t$ be the Index value at time t. B is the set of all values of a Bias, and Bt is the value of the B at time t. n is the number of periods. The foregoing theorem shows that these indices are "matching" indices because any change in the Index value $I_t$ always creates a change in the Bias $B_t$. Then for any time interval t+i, and in any market condition, given the formulas for the Indices, any change $\Delta I$ causes a change $\Delta B$, and B is recursive.
Given the formulas for the Indices, it follows that:
$B_{(t+1)} = F(I_{(t+1)}) = q*F(I_t)$ (the Bias is a function of the Index value).
$\partial|\Delta B_t|/\partial|\Delta I_t| > 0$
$B_{(t+1)} = \Delta B_t + \Delta B_{(t-1)} + ...... \Delta B_{(t-n)}$. (the value of the Bias in any period is equal to the cumulative sum of all the periodic changes in the Bias in prior periods).
Where:
n = the number of prior periods for which the index was calculated.
q = a variable whose value depends on the magnitude of changes in prices of assets in the index. □

*Definition-1*: Noise is defined as any change in an Index (I) within the time interval t+n, that does not reflect the change in the Market ("M"; defined as the value of assets in the market) during t+n.

***Theorem-11: The Above-Mentioned Indices Are Very Noisy.***
***Proof***: Amenc, Goltz, Martellini & Retkowsky (2010), Amenc, Goltz & Le Sourd (Sept. 2006), Choueifaty & Coignard (2008), Clarke, de Silva & Thorley (2006), Fernholz, Garvy & Hannon (1998), Maillard, Roncalli & Teiletche (2008), Martellini (2008), Qian (Sept. 2005), and Qian (2006) describe the formulas for calculating the above mentioned indices. Noise is defined as any change in an Index (I) within the time interval *t+n*, that does not reflect the change in the Market M during *t+n*. To the extent that these Indices are intended to reflect the broad market, for the various reasons and Biases stated above including failure of CAPM/ICAPM, the indices dont reflect the broad market, and hence contain substantial noise. □

***Theorem-12*: *For Any Regular Exchange-Traded Index Futures Contract, Its Price Trend Can Diverge Substantially From Price Trends Of The Underlying Index And Or Cash-market Equivalents Under Some Conditions.***

***Proof***: The price trend of an Index Futures Contract can diverge substantially from price trends of the underlying Index and or cash-market equivalent under any of the following circumstances: **i)** an inverted or very steep yield curve; **ii)** high carrying costs – for debt (interest; transaction costs), and equity (interest; transaction costs), commodities (interest; storage/transportation; transaction costs) and currency (interest); **iii)** the combination of supply-demand imbalance in the Index Units, and announcement/dissemination of new information; **iv)** the cash markets may incorporate new information into prices at a slower rate than the futures markets – due to transactions costs, information asymmetry, availability of capital, beliefs, etc.; **v)** the ETFs that track indices have become quasi-indices themselves that can substantially affect the supply-demand balance for Index units, Index Futures, Index Options and the underlying cash-market securities, all of which may result in substantial differences in price trends of Index futures and the underlying Index; **vi)** ETFs are subject to re-balancing risk and ETF arbitrage around the index re-set dates, which may result in substantial differences in price trends of the Index futures contract and the underlying Index; **vii**) substantial information asymmetry and differences in access to capital between more sophisticated institutional investors and individual investors may also result in substantial differences in price trends of Index futures and the underlying Index; **viii**) the well documented *Ratings Lag* (there is usually a time lag of 6-18 months before credit ratings agencies amend ratings to reflect adverse changes in the financial condition of a company) affects the relationships between futures contracts and the underlying cash markets in equity and debt markets; **ix)** there is substantial index arbitrage that substantially increases or reduces volatility of the cash market; **x)** the differences in the contract sizes (futures contracts are typically much larger), liquidity (futures contracts are more liquid), complexity (futures are used by more sophisticated investors and in more complex spread/arbitrage trades) and the settlement procedures (eg. cash versus delivery) of futures contracts and Index Units and ETF shares stratifies trading activities in these three segments and increases the disconnect (tracking error) between the Index futures, underlying Index and cash market equivalent; **xi)** the differences in cash holdings of participants in the futures markets, cash sub-market and ETF sub-market also stratifies trading activities and investor sentiments that makes Index Futures contracts much less predictive with respect to the underlying Index and cash market equivalent; **xii)** short term (and or medium term) interest rates are lower than the inflation rate (negative real returns); **xiii)** when the compounding of interest has a greater effect on assets in the Index, than on the Index Futures contract, then the Index Futures contract is more likely to diverge from the underlying Index and or cash market equivalent; **xiv)** when the assets in the Index have either very low or very high sensitivity to changes in benchmark interest rates, then the Index Futures contract is more likely to diverge from the underlying Index and or cash market equivalent; **xv)** when during the "roll period" (which is typically announced) arbitrageurs drive up the prices of futures contracts, and thus cause increased divergences among the prices of the Index futures contract and the underlying index and the cash-market eqivalent; **xvi)** when there are substantial differences in quality in the underlying asset (eg. the quality of copper and other physical commodities around the world), its more probable that the Index futures prices will diverge from the underlying Index or cash-market equivalent; **xvii**) when there is substantial de-centralization of physical trade and or exchange-based trading activities (eg. in commodities) the Index futures prices will be more likely to diverge from the underlying Index or cash-market equivalent; **xviii**) a combination of reduced money supply and low liquidity in cash markets is likely to cause increased divergences among the Index futures prices, the underlying Index and or the cash-market equivalent. Aulerich, Fisher & Harris (2011) attempts to explain why expiring futures and cash prices diverge in grain markets. □

***Theorem-13: For Any Futures Portfolio In Which Both The Number Of Futures Contracts And The Absolute Price of Each Constituent Futures Contract Are Un-Restricted, Its Statistically Impossible To Maintain A Constant Portfolio Volatility By Re-Balancing The Futures Contracts And Cash In The Portfolio.***
***Proof***: This is because:

For a two-asset portfolio, the portfolio variance formula is:
$\sigma^2 = X_1^2 \sigma_1^2 + X_2^2 \sigma_2^2 + 2(X_1 X_2 \rho_{12} \sigma_1 \sigma_2)$; and for the general case of N assets, the formula for portfolio variance is:

$$\sigma^2 = \sum_{i=1}^{N} \sum_{j=1}^{N} x_i x_j \sigma_{ij}$$

Where $x_i$ is the weight of asset i in the portfolio. In all instances, the portfolio volatility is the standard deviation, which is the square root of the variance.

**i**) The formula for calculation of portfolio volatility implicitly assumes that there is stationarity of means, variances and covariances which almost never occurs in financial markets (especially for very liquid assets such as exchange traded futures). Even if its assumed that there is instantenuous and continuous portfolio re-balancing, portfolio volatility will never be constant because of transaction costs, taxes, effects of compounding of interest on cash in the portfolio, and the "alternate" volatility of cash (described above).

**ii**) As discussed in this document, most if not all of the formulas for calculating implied volatility and realized volatility are wrong; and historical volatility has been shown to be inaccurate and un-reliable because of entropy.

**iii**) Because the number of Futures contracts in each portfolio (ie. each FTSE StableRisk Index) varies and is not restricted, given the formula for Portfolio Volatility and all else held constant, the greater the number of futures contracts included in the portfolio (ie. the FTSE StableRisk Index), the greater the likelihood that the Portfolio Volatility will converge within a band/range over time. This is a *Population Illusion*, because its not related to the true risk of the constituent futures contracts, but occurs solely because of the number of futures contracts that are used in constructing the portfolio (ie. the FTSE StableRisk Index). Thus, while the calculated Portfolio Volatility may appear constant, in reality, it is not constant.

**iv**) Similarly, because the prices of the individual Futures contracts in each portfolio (ie. FTSE StableRisk Index) vary and are not subject to caps or floors, given the formula for portfolio volatility (which is partly weighted by price) and all else held constant, the greater the differences among the absolute-prices and historical-means-of-prices of selected futures contracts included in the Portfolio (index), the greater the likelihood that the portfolio volatility will be outside a band/range over time. That is, a six-asset futures portfolio $F_1$ (with the following asset prices: 2, 2.5, 3, 2.75, 2.25 and 2.6, and an average volatility of 20%) will have a different portfolio volatility pattern than another similar six-asset futures portfolio $F_2$ (with the following asset prices 1,6, 8,3,10,4 and 2, and an average volatility of 20%). This is an *Absolute-Price Illusion*, because its not related to the true risk of the futures contracts, but occurs solely because of the absolute differences among the absolute-prices of the futures contracts that are used in the portfolio/Index.

**v**) Since trading/transaction costs are deducted when computing portfolio returns for the portfolio (ie. the FTSE StableRisk Index), the amounts available to be invested in cash or futures contracts will vary and this will distort the Portfolio Volatility. □

## 3.11. The Dow Jones RPB Indices.

The Dow Jones RPB Indices (http://www.djindexes.com/rbp/) are one of the very few indices that attempt to include operating performance in index methodologies. However, the RPB indices are wrong because i) they focus on narrow aspects of operations such as sales revenues; ii) the actual index series consist of probabilities; and the process of transformation of operating data into probabilities results in loss of information (partly because of misfit of probability distributions).

**3.12. The FTSE StableRisk Index Series**

The FTSE StableRisk Index Series were created by FTSE Group and AlphaSimplex Group (founded by Prof. Andrew Lo of MIT, Massachusetts, USA). FTSE (2011). Chafkin, Lo & Sinnott (2011). The main stated objective of the FTSE StableRisk Index Series is to reduce divergences between investors' expectations and realized risk levels by capturing long-term expected return premia with less extreme shifts in short-term risk levels which is to be achieved by maintaining constant volatility in any market scenario. Each of four base StableRisk indices (for equity, debt, currency and commodities) is constructed with a portfolio of long-only futures contracts (and cash) which are re-balanced as frequently as daily; while a composite/aggregate Index that supposedly reflects all asset classes is constructed with a portfolio of both long and short futures contracts (and cash). In the FTSE StableRisk Indices, Capital not required for margin is assumed to be held as cash that earns interest based on current money market rates; and Cash returns are simulated as the 1-month LIBOR rate on 80% of the assumed portfolio value, and added to the index value. Thus, the FTSE StableRisk Indices have similar weaknesses as the HSRAI and SPRCI which are discussed in preceding sections. The FTSE StableRisk Indices are inaccurate for the following reasons.

When calculating FTSE StableRisk Indices, Re-balancing occurs only if position changes are at least twenty five percent larger or smaller than the previous position; and this results in more variable Portfolio volatility, lower turnover and transaction costs and greater inaccuracy (the inaccuracy increases as the stated target Volatility declines). The selection of the 25% cut-off point is not based on any reasonable rationale. While Chafkin, Lo & Sinnott (2011) use the Sharpe Ratio to justify the FTSE StableRisk Indices, as shown above, the Sharpe Ratio is wrong.

The construction of FTSE Stable Risk Indices involves the use of a type of equal-risk-weighting wherein within each asset class, risk is allocated equally among countries, and within each country, risk is allocated equally among all constituent contracts; and where countries are not relevant (eg. commodities), risk is allocated equally among all constituent assets. The weaknesses and inaccuracy of equal-risk weighting methods are explained above.

As shown above, the price trend of Index Futures Contracts can diverge substantially from price trends of the underlying Index and or cash-market equivalent. Some empirical and theoretical studies have also confirmed this. The Futures Contracts used in the FTSE StableRisk Indices are based on underlying indices which are inaccurate (don't track the target market accurately) and inefficient as described in Nwogugu (2010c). Aulerich, Fishe & Harris (2011) found that in recent years, cash and futures prices didn't converge at expiration for selected corn, soybean, and wheat commodity contracts; which indicates that arbitrage activities may not have been effective, and futures contracts aren't always useful for hedging. Aulerich, Fishe & Harris (2011) attempted to show that the delivery process for these contracts contain a valuable real option on the long side—the option to exchange the deliverable for another futures contract. As the relative volatility of cash and futures prices increases, this option increases in value, which disconnects the cash market from the deliverable instrument in a futures contract. Tzang, Hung, Wang & Shyu (2011) found that liquidity and sampling methods affect the construction of Volatility Indices. Thus, the selection of futures contracts whose expiration dates don't match the re-set dates and or dividend dates of indices can affect the accuracy and performance of the FTSE StableRisk Indices. The selection of standard Futures contracts instead of the E-Mini futures contracts for use in constructing indices also affects the accuracy of such indices – different classes of traders use different types of futures to obtain exposure to each asset class.

As confirmed by several studies, most estimates of implied volatility and Realized volatility are inaccurate; and thus the FTSE stable-Risk Indices are inaccurate. Szakmary, Ors, Kim & Davidson (2003) used data from thirty five futures-options markets from eight separate exchanges to test whether the implied volatilities (IVs) embedded in option prices predict subsequently realized volatility (RV) in the underlying futures. Szakmary, Ors, Kim & Davidson (2003) found that implied Volatilities don't perform well, but generally outperform historical-volatility (HV) as a predictor of the subsequently RV in the underlying futures prices over the remaining life of the option. Szakmary, Ors, Kim & Davidson (2003) found that in most of the markets they studied, HV didn't contain any economically significant predictive information beyond what was already incorporated in IV regardless of whether it was modeled as a simple moving average or in a GARCH framework. Szakmary, Ors, Kim & Davidson (2003) concluded that their results were consistent with the hypothesis that futures options markets in general, with their minimal trading frictions, are efficient. Konstantinidi & Skiadopoulos (2011) analyzed whether volatility of futures prices can be forecasted by using the VIX futures market; and tested various model specifications (point and interval out-of-sample forecasts were constructed and evaluated under various statistical metrics); and they found only weak evidence of statistically predictable patterns in the evolution of volatility futures prices; and none of the trading strategy they tested yielded economically significant profits. Hence, VIX futures prices cannot be predicted with any significant accuracy, and the hypothesis that the VIX volatility futures market is informationally efficient cannot be rejected. Neely (2009) stated that research has consistently found that implied volatility is a conditionally biased predictor of realized volatility across asset markets. Neely (2009) found that several recently proposed solutions (such as a model of priced volatility risk) fail to explain a significant portion of the conditional bias found in implied volatility; and implied volatility forecasts do not significantly improve delta-hedging performance. Neely (2009) concluded that statistical metrics are inappropriate measures of the information content of implied volatility; and that Implied volatility appears much more useful when measured by a more relevant economic metric. Chang, Hsei & Lai (2009) analyzed the information content of options trading and the predictive power of the put and call positions of different types of traders in the TAIEX option market; and found that options volume has no information on TAIEX spot index changes; and that although foreign institutional investors do not engage in much trading, their trading activities have significant predictive power about the underlying asset returns, and this predictive power is more evident in near-the-money and middle-horizon options. Klitgaard & Weir (May 2004) examined the "net positions of speculators in the futures markets" (currency trading data from the Chicago Mercantile Exchange) from 1993 until 2003 and found that net positions do not appear to be useful for anticipating exchange rate moves during the following week (but the analysis of net positions in a given week provides a 75% chance of correctly predicting exchange rate moves during that same week). CXO Advisory Group(April 2008) conducted an empirical study of the S&P 500 Futures markets and found that *a*ggregate S&P 500 index futures positions (categorized by types of traders, as reported in weekly "Commitments of Traders" reports), may in recent years have lost any significant ability to forecast stock market patterns. The CXO Advisory Group (April 2008 - CXO Advisory Group - http://www.cxoadvisory.com/commodity-futures/extinction-of-the-predictive-power-of-futures/) Study segmented market traders into three categories - commercial hedgers; non-commercial traders (large speculators); and, non-reportable traders (small or retail speculators) and used the data in the weekly "*Commitment of Traders Report*" (published by the Commodity Futures Trading Commission between March 1995 and March 2008 - a total of 675 weeks) as samples.

Futures markets attracts a greater than normal amount of pure speculators. Several researchers have shown that futures markets are subject to manipulation, which often affects the underlying cash markets. Perdue (1987). Futures markets are also highly subject to policy-spillover effects (wherein the market policies of one domestic exchange affect trading and price discovery in foreign exchanges). To the extent that the FTSE StableRisk Indices don't address policy-spillover effects, such indices are inaccurate. Chng (2004) found substantial policy spill-over effects between the Osaka Stock Exchange and the Singapore stock exchange.

It's well documented that the relatively high liquidity and leverage of Index Futures Contracts, Interest rate futures, commodity futures and currency Futures cause increased volatility of the underlying assets and Indices – but the structure of the FTSE StableRisk Indices does not eliminate this adverse symbiotic causal effect. There is also documented evidence that futures contracts respond not only to investor sentiment but also to interest rates, transaction costs, politics and other factors. Given that Futures Contracts (both regular and e-Mini contracts) and the FTSE StableRisk Indices don't isolate investor sentiment, the Indices are inaccurate.

Also, since Volatility does not measure directional trend (upwards or downwards trends), it cannot be used to determine or monitor risk in this context – that is, maintaining a constant volatility is insufficient to hedge market movements. Nwogugu (2003; 2005; 2010a), Becker, Clements & White (2007); Bollen & Whaley (2004); Corrado & Miller (2005); Dotsis, Psychoyios & Skiadopoulos (2007), Taleb (2008) and Giot (2005b) have all documented the major problems inherent in using volatility (historical volatility; implied volatility; or realized volatility) as a measure of risk. It has been shown that other factors distinct from intrinsic risk affect Realized Volatility and Implied Volatility and such factors include – liquidity, buying/selling pressure; availability of capital, etc.. Nossman & Wilhelmsson (2009) tested the expectation hypothesis by analyzing VIX and VIX futures contracts; and found that the VIX futures price contained a negative risk premium, and that when the futures price is not adjusted with the risk premium, the expectation hypothesis is rejected at the 5% significance level for 20 of 21 forecast horizons. They also found that when the futures price is adjusted with the risk premium (derived from a stochastic volatility model), the expectation hypothesis cannot be rejected; and the risk premium adjusted futures price forecasts the direction of the VIX index well (the one–day-ahead forecast predicts the direction correctly 73% of the time). However, the error in this study is the adjustment of the futures price – most stochastic volatility models are inaccurate.

Given the foregoing, the FTSE StableRisk Indices have not solved the problem of Stationarity (the incorrect assumption that means, variances and correlations are constant), and its clear that futures contracts in equities, debt, currencies and commodities markets are not accurate predictors of the trends in the underlying markets, and that Volatility (implied or historical) is not a good measure of risk; and thus, the FTSE StableRisk Indices are inaccurate.

## Conclusion

Indexing has been a major element of asset management (more than US$10 Trillion of capital is managed in, or committed to Index funds in the US alone). The existing problems inherent in index calculation methods arise from various sources including the structure of Indices, investors' reactions to changes in indices, mis-understanding of inaccurate but generally-accepted finance theories, inadequate regulation; adverse effects of Index Arbitrage, concerns about transactions costs, taxation, etc.. The major problems are that the true properties of these Indices differ from what they are marketed as, and some of the Index sponsors earn sales commissions from sales of Index funds – all of which implies substantial and adverse information asymmetry and moral hazard. Furthermore, the existing index calculation methods dont capture or aggregate investors' preferences or the true economic "footprint" of companies in the economy or the true operational risk of companies. Although many regulators and investors erroneously use Indices as measures of conditions and trends in the economy, there is a significant dichotomy and divide between current values of Indices and economic trends in the economies that the Indices are supposed to reflect – this has substantial policy implications. The existing risk-weighted index calculation methods erroneously assume full observability of investors' preferences, whereas securities investments typically account for a relatively small percentage of the Total Wealth of the average investor. In these Risk-weighted index calculation methods, risk is defined primarily within the Mean Variance Framework, and emphasizes returns and standard deviations, whereas investor preferences are defined by a much broader set of metrics. Also, some of the above-mentioned index weighting methods include static Index Revision Dates which have adverse effects and facilitate harmful Index arbitrage. Existing Risk-weighted index calculation methods increase inherent risks of, and reduce the "financial stability" of markets because they: i) propagate market noise that is un-related to the fundamental performances or to the operational risk of companies in the Indices, ii) propagate a focus on variances which creates volatility which in turn requires more hedging and thus creates more volatility, iii) focus on a few metrics that are insufficient to define relative risk of assets and un-biased trends in markets. Thus, current Risk-weighted Index calculation methods have substantial adverse effects on social welfare because of the size, scope and effects of the Index funds market; and there is a significant need for new Index calculation methods that reduce or eliminate the problems and Biases described above.

# **Bibliography**

**1.** Aksu, M.H. & Onder T. (2003). *The Size and Book-to-Market Effects and Their Role as Risk Proxies in the Istanbul Stock Exchange.* SSRN Working Paper.

**2.** Alexander C & Barbosa A (2007). Effectiveness of Minimum-Variance Hedging: *The impact of electronic trading and exchange-traded funds. Journal of Portfolio Mangement*, pp 1-7.

**3.** Amenc N, Goltz F, Martellini L & Retkowsky A (2010). Efficient Indexation: An Alternative to Cap-Weighted Indices. *Journal Of Investment Management*, ____________.

**4.** Amenc, N., Goltz F & Le Sourd V (Sept. 2006). *Assessing the Quality of Stock Market Indices: Requirements for Asset Allocation and Performance Measurement*. http://www.edhec-risk.com/indexes/indices_study/index_html/attachments/Af2i_EDHEC_indices_study.pdf. EDHEC Publication.

**5.** Amenc N, Goltz F, Martellini L & Ye S (2011). Improved Beta? A Comparison Of Index-Weighting Schemes. *Journal of Indexes*, ________.

**6.** Arnott R, Kalesnik V, Moghtader P & Scholl (Jan/Feb. 2010). Beyond Cap Weight: The empirical evidence for a diversified beta. *Journal Of Indexes*, pp. 16-26.

**7.** Aulerich, N. M., Fishe, R. P. H. & Harris, J. H. (2011). Why do expiring futures and cash prices diverge for grain markets?. *Journal of Futures Markets*, 31: 503–533.

**8.** Aydin S & Ozer G (2005). National customer satisfaction indices: an implementation in the Turkish mobile telephone market. *Marketing Intelligence & Planning*, 23(5): 486 – 504.

**9**. Backus D, Routledge B & Zin S (December 2005). Recursive Preferences. NYU Working Paper No. EC-05-19. Available at SSRN: http://ssrn.com/abstract=1282544

**10.** Berger D & Pukthuanthong K (2012). Market fragility and international market crashes. *Journal of Financial Economics*, _______.

**11.** Berk, J. B. (1995). A Critique of Size-Related Anomalies. *Review of Financial Studies*, 8: 275-286.

**12.** Bernhard R (1971). A Comprehensive Comparison and Critique of Discounting Indices Proposed For Capital Investment Evaluation. *The Engineering Economist*, 16(3): 157 – 186.

**13.** Bi G, Ding J, Luo Y & Liang L (2011). A New Malmquist Productivity Index Based On Semi-Discretionary Variables With An Application To Commercial Banks of China. *International Journal of Information technology And Decision Making*, 10: 713-717.

**14.** Blitzer D (Standard & Poors) (2004). *Free Float Adjustment For The S&P*.

**15.** Blume, M. E., & Stambaugh R F (1983). Biases in Computed Returns: An Application to the Size Effect. *Journal of Financial Economics*, 12: 387-404.

**16.** Caves D, Christensen L & Diewert R (1982). The Economic Theory Of Index Numbers And the Measurement of Input, Output And Productivity. *Econometrica*, 50(6): 1393-1414.

**17.** Chafkin J H, Lo A& Sinnott R (2011). The FTSE StableRisk Indices. Available at: http://www.ftse.com/Indices/FTSE_StableRisk_Composite_Trend_Index_Series/Downloads/FTSE_StableRisk_Indices_Whitepaper_May_2011.pdf.

**18.** Chan, L. K. C., Karceski J & Lakonishok J. (1999). On portfolio optimization: Forecasting Covariances And choosing the risk model. *Review of Financial Studies*, 12(5): 937-74.

**19.** Chen T (2012). Non Linear Assignment-Based Methods For Interval-Valued Intuitionistic Fuzzy Multi-Criteria Decision Analysis With Incomplete Preference Information. *International Journal of Information Technology & Decision Making*, 11: 821-827.

**20.** Chng M (2004). The trading dynamics of close-substitute futures markets: evidence of margin policy spillover effects. *Journal of Multinational Financial Management*, 14(4-5): 463-483.

**21.** Choueifaty, Y., & Coignard Y (2008). Toward Maximum Diversification. *Journal of Portfolio Management*, 35(1): 40-51.

**22.** Clarke R., de Silva H & Thorley S. (2006). Minimum-variance portfolios in the U.S. Equity Market. *Journal of Portfolio Management* (fall): 1-14.

**23**. Conrad, J. & Kaul G (1993). Long-Term Market Overreaction or Biases in Computed Returns. *Journal of Finance*, 48(1): 39-63.

**24.** CXO Advisory Group (April 2008). *Extinction of the Predictive Power of Futures* ? Available at: http://www.cxoadvisory.com/commodity-futures/extinction-of-the-predictive-power-of-futures/.
**25.** Daniel, K. & Titman, S. (1997). Evidence On The Characteristics Of Cross Sectional Variation In Stock Returns. *Journal of Finance*, 52: 1–33.
**26.** Danielsson J, Jorgensen B & Sarma M & Vries C (2006). Comparing downside risk measures for heavy tailed distributions. *Economic Letters*, 92(2): 202-208.
**27.** DeMiguel, V., Garlappi L & Uppal, R (2007). Optimal Versus Naive Diversification: How Inefficient Is the 1/N Portfolio Strategy? *Review of Financial Studies*, ________.
**28.** DesHarnais, S.I., Forthman, M.T., Homa-Lowry, J.M. & Wooster, L.D. (2000). Risk-Adjusted Clinical Quality Indicators: Indices for Measuring and Monitoring Rates of Mortality, Complications, and Readmissions. *Quality Management in Health Care*, 9(1): 14-22.
**29.** DesHarnais, S. I., Forthman, M.T., Homa-Lowry, J.M. & Wooster, L.D. (1997). How to Use Risk-Adjusted Quality Indicators to Assess Hospitals. *QRC Advisor*, 13(5).
**30.** Diewert E (2009). Cost of Living Indexes and Exact Index Numbers Revised. In Slottje D, ed., (2009), "*Quantifying Consumer Preferences*" (Contributions to Economic Analysis, Volume 288) (Emerald Publishing Group, 2009).
**31**. Dorfleitner, G (2003). Why the return notion matters. *International Journal of Theoretical & Applied Finance*, 6: 73–86.
**32.** EDHEC Risk Institute (Oct. 2010). METHODOLOGY FOR THE MANAGEMENT OF THE FTSE EDHEC-RISK EFFICIENT INDEX SERIES. Available at: http://docs.edhec-risk.com/nwl/110131/FTSE_EDHEC-Risk_Efficient_Index_Series_Rules.pdf.
**33.** Elton E. J., Gruber, M. J. & Busse J. A. (2004). Are investors rational? Choices among index funds. *Journal of Finance*, 59: 261–88.
**34**. Epaulard A & Pommeret A (Jan. 2001). *Recursive Utility, Endogenous Growth And The Welfare Cost of Volatility*. IMF Working Paper WP/01/5.
**35.** Falkenstein, E G. (2009). *Risk and Return in General: Theory and Evidence*. Available at SSRN: http://ssrn.com/abstract=1420356
**36.** Fan G & Zeng Y (2012). The Timing Of portfolio Adjustments: A Regime Switching Approach. *International Journal Of Information technology And Decision Making*, 11: 909-914.
**37.** Fernholz, R., Garvy R & Hannon J (1998). Diversity-Weighted Indexing. *Journal of Portfolio Management*, 4(2): _______.
**38**. Fisher, L., Weaver D & Webb G (June 2009). *Removing Biases in Computed Returns*. Working paper, Rutgers University.
**39**. Fisher L., Weaver D. & Webb G. (______). Removing Biases in Computed Returns: An Analysis of Bias in Equally-Weighted Return Indexes of REITs. http://asres2008.shufe.edu.cn/session/papers/d72.pdf.
**40.** Flam S (2010). Portfolio Management Without Probabilities or Statistics. *Annals Of Finance*, _______.
**41.** Forthman, M.T., Gold, R.S., Dove, H.G. & Henderson, R.D. (2010). Risk-Adjusted Indices for Measuring the Quality of Inpatient Care. *Quality Management in Health Care*, 19(3): 265-277.
**42.** Frino A, Gallagher D R & Oetomo T N (2005). The Index Tracking Strategies of Passive and Enhanced Index Equity Funds. *Australian Journal of Management*, 30: 23 - 55.
**43.** Garlappi L, Shu T & Yan H (2008). Default Risk, Shareholder Advantage, and Stock Returns. Review Of Financial Studies, 21(6): _______.
**44.** Gharghori P, Chan H & Faff R (2007). Are The Fama-French Factors Proxying Default Risk? *Australian Journal of Management*, 32: 223 - 249.
**45.** Green R & Hollifield B (1992). When Will Mean-Variance Portfolios be Well Diversified ?. *Journal of Finance*, 47: ________.
**46**. Guo H (May/June 2004). *A Rational Pricing Explanation For The Failure Of The CAPM*. Federal Reserve Bank of St. Louis, pp. _________.
**47.** Hertzberg, M. (1987). *A Critique Of The Dust Explosibility Index: An Alternative For Estimating Explosion Probabilities*. U.S. Department of the Interior, Bureau of Mines, Report of Investigations, RI 9095.
**48.** Hsu, J. (2006). Cap-Weighted Portfolios are Sub-optimal Portfolios. *Journal of Investment Management*, 4(3): 1–10.

**49.** Hoque A. (Jan. 2010). *Econometric Modeling for Transaction Cost-Adjusted Put-Call Parity: Evidence from the Currency Options Market*. Available at SSRN: http://ssrn.com/abstract=1537834.
**50.** Hsu, J C. & Campollo, C (Feb. 6, 2006). New Frontiers In Index Investing. *Journal of Indexes,* January/February 2006.
**51.** Hurlbert S (1971). The Non-Concept Of Species Diversity: A Critique And Alternative parameters. *Ecology*, 52(4) 576-580.
**52.** Haugen, R.A., and Baker, N. L. (1991). The Efficient Market Inefficiency of Capitalization-Weighted Stock Portfolios. *Journal of Portfolio Management*, 17(3): 35–40
**53.** Jha, R & Murthy K. & Bhanu V (April 2003). *A Critique of the Environmental Sustainability Index*. Australian National University Division of Economics Working Paper. Available at SSRN: http://ssrn.com/abstract=380160 or doi:10.2139/ssrn.380160.
**54.** Jianping L. et al (2012). Risk Integration Mechanisms And Approaches In The Banking Industry. *International Journal of Information technology And Decision Making,* 11: 1183-1187.
**55.** Joyce J & Vogel R (1970). The Uncertainty In Risk: Is Variance Unambiguous ? *Journal Of Finance,* 25(1): 127-134.
**56.** Karabatsos G (2000). A Critique Of Rasch Residual Fit Statistics. *Journal Of Applied Measurement*, 1(2): 152-176. http://tigger.uic.edu/~georgek/HomePage/Karabatsos2000JAM.pdf.
**57.** Keim, D. B. (1999). An analysis of mutual fund design: The case of investing in small-cap stocks. *Journal of Financial Economics*, 51: 173–94.
**58.** Kim J, Kim S H & Levin A (2003). Patience, Persistence And Welfare Costs Of Incomplete Markets In Open Economies. *Journal Of International Economics*, 61(2): 385-396.
**59**. Klibanoff P, Marinacci M & Mukerji S (2009). Recursive smooth ambiguity preferences**.** *Journal of Economic Theory*, 144(3): 930-976.
**60.** Klitgaard & Weir (May 2004). Exchange Rate Changes And Net Positions Of Speculators In Futures Markets. *Economic Policy Review – Federal Reserve Bank Of New York*, 10(1).
**61.** Konstantinidi E & Skiadopoulos G (2011). Are VIX futures prices predictable? An empirical investigation. *International Journal of Forecasting*, 27(2): 543-560.
**62.** Konstantinidi E., Skiadopoulos, G & Tzagkaraki, E (2008). Can the evolution of implied volatility be Forecasted ? Evidence from European and US implied volatility indices. *Journal of Banking and Finance*, 32(11): 2401-2411.
**63.** Kumar V & Ziemba W (1993). The Effect of Errors in Means, Variances, and Covariances on Optimal Portfolio Choice. *Journal of Portfolio Management*, 19(2): 6-11.
**64.** Lewellen, J. & Nagel S (2006). The Conditional CAPM Does Not Explain Asset-Pricing Anomalies. *Journal of Financial Economics*, 82: 289-314.
**65.** Li D (2009). Relative Ratio Method for Multiple Attribute Decision Making Problems. *International Journal of Information Technology & Decision Making*, 8(2): 289–311. **58.** Loayza N, Ranciere R, Serven L & Ventura J (2007). Macroeconomic Volatility and Welfare In Developing Countries: An Introduction. *World Bank Economic Review*, 21(3): 343-357.
**66.** Madhavan A. & Ming, K. (2002). *The hidden costs of index re-balancing: A case study of the S&P 500 composition changes of July 19, 2002*. ITG Working Paper.
**67.** Maillard S., Roncalli T & Teiletche J. (2008). *On The Properties Of Equally-weighted Risk Contributions Portfolios*. Working paper.
**68.** Mar J, Bird R, Casavecchia L & Yeung D (2009). Fundamental Indexation: An Australian Investigation. *Australian Journal of Management*, 34: 1-20. http://www.agsm.edu.au/eajm/0906/pdf/Paper1-0906.pdf.
**69.** Martellini L (2008). Towards the Design of Better Equity Benchmarks: Rehabilitating the Tangency Portfolio from Modern Portfolio Theory. *Journal of Portfolio Management*, 34(4).
**70.** Morris, S (Dec. 12, 2006). "*What's the Right Way To Index*?". http://news.morningstar.com/article/pfarticle.asp?keyword=indexfunds&pfsection=Index.
**71**. MSCI (May 2011). *MSCI Factor Indices Methodology*. Available at: http://www.msci.com/eqb/methodology/meth_docs/MSCI_Factor_Index_Methodology_May11.pdf.
**72**. MSCI Barra (Nov. 2009). *MSCI Minimum Volatility Indices Methodology*. Available at: http://www.mscibarra.com/products/indices/thematic_and_strategy/minimum_volatility/MSCI_Minimum_Volatility_Methodology.pdf

**73.** Murphy E & Garvey E. (2005). *Cost of Living Indices and Flexible Consumption Behaviour: A partial critique*. Working Paper.
**74.** Neely C J (2009). Forecasting foreign exchange volatility: Why is implied volatility biased and inefficient? And does it matter?. *Journal of International Financial Markets, Institutions and Money*, 19(1): 188-205.
**75.** Neher D & Darby B (______). *Computation and application of nematode community indices: general guidelines*. http://www.uvm.edu/~dneher/Publications/04Neher_Darby%20chapter.pdf.
**76.** Nwogugu M. (2003). Decision-Making Under Uncertainty: A Critique Of Options Pricing Models. *Derivatives Use, Trading And Regulation* (now *Journal Of Derivatives & Hedge Funds*), 9(2): 164-178.
**77.** Nwogugu M. (2005a). Further Critique Of GARCH/ARMA/VAR/SV Models. *Applied Mathematics & Computation*, 182(2): 1735-1748.
**78.** Nwogugu M. (2005b). Towards Multifactor Models Of Decision Making And Risk: Critique Of Prospect Theory And Related Approaches, Part Three. *Journal Of Risk Finance*, 6(3): 267-276.
**79.** Nwogugu M (2011). *Eliminating Index Arbitrage And ETF Arbitrage: Non-Legislative Methods*. Working Paper.
**80.** Nwogugu M (2010a). Correlation, Variance, Co-Variance And Semi-variance Are Irrelevant In Porfolio Management. Working Paper.
**81.** Nwogugu M (2010b). *CML, ICAPM/CAPM and APT/IAPT Are Inaccurate In Incomplete Markets With Dynamic Un-Aggregated Preferences*. Working Paper.
**82**. Nwogugu M (2010c). *Recursive "Matching" Noise And Biases In Traditional Index Calculation Methods In Incomplete Markets With Un-Aggregated Preferences*. Working Paper- available at www.ssrn.com.
**83.** Nwogugu M (2010d). *Some Biases In The Calculation Of Returns*. Working Paper, www.ssrn.com.
**84**. Pallage S & Robe M (2003). On The Welfare Cost Of Economic Fluctuations In Developing Countries. *International Economic Review*, 44(2): 677-698.
**85.** Powell P, Roa R., Shi J. & Xayavong V. (Dec. 2007). A Test for Long-Term Cyclical Clustering of Stock Market Regimes. *Australian Journal of Management*, 32: 205 - 221.
**86**. Prono T (June 2009). *Market Proxies, Correlation And Relative Mean-Variance Efficiency: Still Living With The Roll Critique*. Working Paper # QAU09-3, Federal Reserve Bank Of Boston, USA.
**87.** Prono T (June 2007). *GARCH-based Identification Of Triangular Systems With An Application To The CAPM: Still Living With The Roll Critique*. Working Papers #07-1, Federal Reserve Bank of Boston. *http://www.bos.frb.org/economic/wp/wp2007/wp0701.pdf.*
**88.** Qian E. (Spet. 2005). *Risk parity portfolios: Efficient portfolios through true diversification*. Panagora Asset Management.
**89.** Qian E. (2006). On the Financial interpretation of risk contributions: Risk budgets do add up. *Journal of Investment Management*, __________.
**90.** Ramsden J J (2009). Impact Factors: A Critique. *Journal Of Biological Physics & Chemistry*, 9: __________.
**91.** Ronalds N & Anderson C (Nov./Dec. 2006). The Synthetic EAFE Index. *Journal Of Indexes*, _________.
**92.** Russell (2008). *US Equity Style Methodology – Comments*.
**93.** Sault S (2005). Movements in Australian Stock Volatility: A Disaggregated Approach. *Australian Journal of Management*, 30: 303 - 320.
**94.** Scherer B (Sept. 2011). A New Look At Minimum Variance Investing. *Journal Of Empirical Finance*, _____________.
**95.** Schultz M T (2001). A Critique Of EPA's Index of Watershed Indicators. *Journal of Environmental Management*, 62(4): 429-442.
**96.** Serrano R & Aumann R J (2008). An Economic Index of Riskiness. " *Journal of Political Economy*, _________________.
**97.** Standard & Poors (August 2011). *S&P Index Mathematics Methodology*.
**98**. Taleb N (2008). Finiteness Of Variance Is Irrelevant In The Practice Of Quantitative Finance. *Complexity*, 14(3): 66-76.
**99.** Su Z (2011). A Hybrid Fuzzy Approach To Fuzzy Multi Attribute Group Decision Making. *International Journal Of Information Technology And Decision Making*, 10: 695-700.
**100.** Szakmary A, Ors E, Kim J & Davidson W (2003). The predictive power of implied volatility: Evidence from thirty five futures markets. *Journal Of Banking & Finance*, 27: 2151–2175.

**101.** Takahashi H (2012). An Analysis of The Influence of Dispersion Of Valuations On Financial Markets Through Agent-Based Modeling. *International Journal Of Information technology & Decision Making*, 11: 143-148.
**102**. Taleb N (2008). Finiteness Of Variance Is Irrelevant In The Practice Of Quantitative Finance. *Complexity* 14(3): 66-76.
**103.** Thomson Reuters(2007). Lipper Optimal Indices. Available at: http://www.lipperweb.com/Handlers/GetDocument.ashx?documentId=4201.
**104.** Tofallis, C (2008). Investment Volatility: A Critique of Standard Beta Estimation and a Simple Way Forward. *European Journal of Operational Research*, 187: 1358-1367.
**105.** Tucker T. (1997). Rethinking Rigor In Calculus: The Role Of The Mean Value Theorem. *American Mathematical Monthly*, 104(3): 231–240.
**106.** Tzang S, Hung C, Wang C & Shyu D (2011). Do Liquidity And Sampling Methods Matter In Constructing Volatility Indices? Empirical Evidence From Taiwan. *International Review of Economics & Finance*, 20(2): 312-324.
**107.** U.S. Department Of Commerce/National Oceanic and Atmospheric Administration (2003). *Report On Wind Chill Temperature and Extreme Heat Indices: Evaluation And Improvement Projects*. Available at: http://www.ofcm.gov/jagti/r19-ti-plan/pdf/00_opening.pdf.
**108.** Wong D (April 2011). *Hang Seng Risk Adjusted Index Series*. Available at: http://www.hsi.com.hk/HSI-Net/static/revamp/contents/en/dl_centre/presentations/P_HSRAIe_20110324.pdf.

**1.** Adler T & Kritzman M (2007). Mean-Variance versus Full-Scale Optimization: In And Out of Sample. *Journal of Asset Management***,** 7:302–311.
**1.** Kroll Y, Levy H & Markowitz H (1984). Mean-Variance Versus Direct utility Maximaization. *Journal of Finance*, 39(1):47-61.
**51.** Martellini L & Urosevic B (2006). Static Mean-Variance Analysis With Uncertain Time Horizon. *Management Science*, _______________.